\documentclass[11pt,a4paper]{article}
\usepackage[T1]{fontenc}
\usepackage[utf8]{inputenc}
\usepackage{authblk}
\usepackage{fullpage}
\usepackage{graphicx}
\usepackage{xcolor}
\usepackage{algorithm}
\usepackage{algpseudocode}
\usepackage{amssymb}
\usepackage{amsthm}
\usepackage{amsmath}
\usepackage{url}
\usepackage{bbold}
\usepackage{stmaryrd}
\usepackage{txfonts}
\usepackage[square,numbers,comma,sort&compress]{natbib} 
\usepackage[colorlinks,urlcolor=red,citecolor=red,linkcolor=blue]{hyperref}
\newcommand{\ditto}{~\textquotedbl~}
\title{Non-uniform FFT for the finite element computation of the micromagnetic scalar potential}
\author[1]{Lukas Exl\thanks{Corresponding author, \texttt{lukas.exl@univie.ac.at}}}
\author[2]{Thomas Schrefl}

\affil[1]{University of Vienna, Fakult\"at f\"ur Mathematik, Oskar-Morgenstern-Platz 1, A-1090 Wien, Austria}
\affil[2]{University of Applied Sciences, Department of Technology, A-3100 St.Poelten, Austria}

\begin{document}
\maketitle
\begin{abstract}
We present a quasi-linearly scaling, first order polynomial finite element method for the solution of the magnetostatic open boundary problem by splitting the magnetic scalar potential. 
The potential is determined by solving a Dirichlet problem and evaluation of the single layer potential by a fast approximation technique based on Fourier approximation of the kernel function. The latter approximation
leads to a generalization of the well-known convolution theorem used in finite difference methods. We address it by a non-uniform FFT approach. Overall, our method scales $\mathcal{O}(M+N+N\log N)$ for $N$ nodes and $M$ surface triangles. 
We confirm our approach by several numerical tests.
\end{abstract}

\begin{center}
\end{center}
\section{Introduction}
In micromagnetic simulations one is interested in either finding \textit{magnetization configurations} of minimal magnetic energy or the time evolution of the magnetization under influence of internal and external fields \cite{fidler_2000}. In both cases the 
effective field, which consists of \textit{stray field}, \textit{anisotropy, exchange, external} and possibly \textit{thermal field} and \textit{spin transfer torque
interaction} (influence of external currents), has to be computed in each step of an iterative procedure \cite{d2005geometrical}. Among them the stray field part is the most time consuming.\\
Since we are interested in developing a new method for the computation of the latter one, we briefly state here the interface problem which defines the scalar potential of the stray field and give an overview of some existing numerical methods which address it.\\  
The micromagnetic \textit{stray field} is given as $\boldsymbol{h}_s = -\nabla \phi$, where the \textit{scalar potential} $\phi$ for a given  magnetization $\boldsymbol{m} \in \big( C^1 (\Omega) \big)^3 \cap \big( C^0 (\overline{\Omega}) \big)^3$ \footnote{$C^q(\Omega)$ is the space of $q$ times continuously differentiable functions defined on $\Omega$; $\overline{\Omega}$ means closure of $\Omega \subset \mathbb{R}^3$ (bounded and open) and $(.)^c$ stands for complement.}, $\Omega \subset \mathbb{R}^3$ bounded and open, fulfills the interface problem \cite{jackson_classical_1999,engel2007handbook} 
\begin{align}\label{scpot_classical}
-\Delta \phi & \, =  -\nabla \cdot \boldsymbol{m} & \text{in} \,\, \Omega, \notag\\
-\Delta \phi & \, = 0 & \text{in} \,\, \overline{\Omega}^c, \notag\\
\left[\phi\right] & \, = 0 & \text{on} \,\, \partial \Omega, \\
\left[\frac{\partial \phi}{\partial \boldsymbol{n}}\right] & \, = - \boldsymbol{m} \cdot \boldsymbol{n} & \text{on} \,\, \partial \Omega, \notag\\
\phi(\boldsymbol{x}) & \, = \mathcal{O}(\tfrac{1}{\left\| \boldsymbol{x} \right\|}) \quad \text{as} \,\, \left\| \boldsymbol{x} \right\| \rightarrow \infty,&\notag
\end{align}
where $\left[.\right]$ stands for the jump at the boundary. The classical solution of \eqref{scpot_classical} has to be determined in whole space and is at least two times continuously differentiable in $\Omega$ and the exterior region $\overline{\Omega}^c$.\\ 
Since we aim to compute a \textit{weak solution} we further assume that $\Omega$ is a Lipschitz domain with polyhedral boundary $\partial\Omega$. One reformulates the set of equations \eqref{scpot_classical}, \cite{aurada2012classical}:\\[0.1cm] 
For given $\boldsymbol{m} \in \big( H^1 (\Omega) \big)^3$ the \textit{micromagnetic scalar potential} $\phi := (\phi^{int}, \phi^{ext}) \in H^1(\Omega) \times H^1_{loc}(\overline{\Omega}^c)$ is the solution to \eqref{scpot_classical}, where
$-\Delta \phi^{int}  =  -\nabla \cdot \boldsymbol{m} \,\, \text{in} \,\, \Omega$ and $-\Delta \phi^{ext} = 0 \,\, \text{in} \,\, \overline{\Omega}^c$ holds in a variational sense. 
Hereby, $H^1(\Omega)$ denotes the usual \textit{Sobolev space}, i.e. $H^1(\Omega) := \{u \in L^2(\Omega) \mid \text{weak derivatives}\,\, \partial_q u \in L^2(\Omega),\, q = x,y,z \}$ and $H^1_{loc}(\overline{\Omega}^c) := \{u \in H^1(C) \mid C \subset \overline{\Omega}^c \, \text{compact}\}$. The jump $\left[.\right]$ is determined by taking the corresponding trace operators.
Within this setting the existence of a unique solution to \eqref{scpot_classical} has been proved. For details we refer the reader to \cite{aurada2012classical,carstensen1995adaptive} and references therein.\\[0.1cm]
For comparing different algorithms we mention the important fact that in micromagnetic methods it is possible to precompute certain quantities that do not depend on the magnetization, e.g. only rely on the geometry of the problem, see for instance \cite{abert_2013}. 
Thus, in our asymptotic operation counts, we neglect the effort for computing these steps and refer to as \textit{precomputation} or \textit{setup phase}.\\
Several methods address the approximation of the solution to \eqref{scpot_classical}. 
Integral methods with a regular discretization of the domain $\Omega$ aim to directly compute the integral representation of the solution inside the magnetic body $\Omega$ \cite{donahue_2007,abert_2013}, i.e.
\begin{align}\label{int_rep}
\phi(\boldsymbol{x}) = -\frac{1}{4\pi} \Big( \int_{\Omega} \frac{\nabla \cdot \boldsymbol{m}(\boldsymbol{y})}{\left\| \boldsymbol{x} - \boldsymbol{y}\right\|} \,\text{d}\boldsymbol{y} +  \int_{\partial\Omega} \frac{\boldsymbol{m}(\boldsymbol{y}) \cdot \boldsymbol{n}(\boldsymbol{y})}{\left\| \boldsymbol{x} - \boldsymbol{y}\right\|} \,\text{d}\sigma(\boldsymbol{y}) \Big).
\end{align}
Discretization on an equispaced grid built of rectangular computational cells allows applying fast Fourier transform (FFT) techniques, making this methods quasi-optimal, i.e. the costs are $\mathcal{O}(N\log N)$ for $N$ grid points \cite{engel2007handbook,donahue_2007,abert_2013}. 
Also the \textit{fast multipole method} and combination with FFT was applied to compute the magnetostatic field and energy \cite{blue_1991,long_2006,van2008application}, as well as a nonuniform grid (NG) algorithm \cite{livshitz,chang}. Within the framework of integral approaches also tensor grid methods were developed, which make further assumptions on the representation of the magnetization field through tensor formats, but then gain even sub-linear complexity \cite{goncharov_2010,exl_fast_2012_2, exl2012fft}.\\
A precorrected-FFT method was introduced in the context of electromagnetic boundary integral equations \cite{phillips}. In the relevant literature, to the authors' best knowledge, this method has not been adapted to the corresponding micromagnetic equations, but so-called non-uniform FFT could be considered as a related approach.\\ 
A method using non-uniform FFT from \cite{keiner2009using} on the quadrature approximation of the integral representation \eqref{int_rep} discretized on unstructured $2$-dimensional FE grids was reported in \cite{kritsikis2008fast}. 
This method scales $\mathcal{O}(Q + N + n^d\,\log n)$ in the general case of $d$ dimensions for $Q = q L$ quadrature points in total, where $q$ quadrature points are used for each of the $L$ computational domains (tetrahedrons for volume and triangles for surface integrals), $N$ mesh-nodes and an auxiliary parameter $n$, which comes from the FFT and is of the same order as in our forthcoming proposed method, see \cite{potts2003fast} for more details. 
We stress that our method leads to the same complexity for a prescribed accuracy but requires only a pre-factor $q = 1$. This is achieved by performing the integration in a setup phase.\\
Moreover, \textit{shell transformation techniques} on a finite element mesh containing $\Omega$ were applied to address unbounded problems like \eqref{scpot_classical}, \cite{brunotte_1992}. 
The discrete formulation of \eqref{scpot_classical} translates to only one sparse linear system, which, however, tends to be very ill-conditioned due to the transformation. Algebraic multigrid preconditioners were successfully applied to address this issue \cite{abert_2013}.\\    
On the other hand, the well-known hybrid FEM-BEM coupling by the \textit{ansatz of Fredkin and Koehler} \cite{fredkin_1990} aims to solve \eqref{scpot_classical} by the splitting $\phi = \phi_1 + \phi_2$, where $\phi_1$ is determined by a Poisson equation with Neumann boundary conditions
and $\phi_2$ by a Laplace equation where the Dirichlet data are computed by the values of $\phi_1$ through a boundary integral representation of the potential $\phi_2$. Hereby, the calculation of the boundary values of $\phi_2$ leads to a dense matrix-vector
product which scales $\mathcal{O}(N_b^2)$ for $N_b$ boundary nodes. Compression techniques were introduced to reduce this complexity and storage requirements \cite{knittel_2009}.\\[0.1cm]
In this work we present a first order polynomial (P$1$) finite element method that solves \eqref{scpot_classical} by the \textit{ansatz of Garc\'{i}a-Cervera and Roma} \cite{garcia2006adaptive}, where we develop a fast evaluation technique for the single layer potential.
Approximation of a smoothed version of the Newtonian kernel $N(x) = 1/|x|$ by a Fourier series will lead us to a computational scheme which is similar to the convolution theorem used e.g. 
in integral methods mentioned above.  Based on FFT for non-equispaced data (non-uniform FFT, NFFT) \cite{dutt1993fast,beylkin1995fast,potts2001fast} and linearly scaling near-field correction, we are able to efficiently compute the single layer potential. 
We then combine this solution with the finite element solution of a Dirichlet problem, yielding, in total, a complexity of $\mathcal{O}(M+N)$ for $N$ nodes and $M$ surface triangles.\\
In the following section we discuss the \textit{ansatz of Garc\'{i}a-Cervera and Roma}, which will be the basis for our method. 
\section{The ansatz of Garc\'{i}a-Cervera and Roma}
In the following we will describe a quasi-linearly scaling method for the \textit{ansatz of Garc\'{i}a-Cervera and Roma} \cite{garcia2006adaptive}. We split the potential into $\phi = \phi_1 + \phi_2$ and get for $\phi_1 = (\phi_1^{int}, \phi_1^{ext}) \in H^1(\Omega) \times H^1_{loc}(\overline{\Omega}^c)$
\begin{align}\label{GR1}
-\Delta \phi_1^{int} & \, =  -\nabla \cdot \boldsymbol{m} & \text{in} \,\, \Omega,\\
\phi_1^{int} & \, = 0 & \text{on} \,\, \partial \Omega, \notag
\end{align}
and set $\phi_1^{ext} = 0$ in $\overline{\Omega}^c$. Hence, we have $\left[\frac{\partial \phi_1}{\partial \boldsymbol{n}}\right] = - \frac{\partial \phi_1^{int}}{\partial \boldsymbol{n}} $.\\
The second part $\phi_2 = (\phi_2^{int}, \phi_2^{ext}) \in H^1(\Omega) \times H^1_{loc}(\overline{\Omega}^c)$ consequently fulfills
\begin{align}\label{GR2}
-\Delta \phi_2^{int} & \, =  0 & \text{in} \,\, \Omega,\notag\\
-\Delta \phi_2^{ext} & \, =  0 & \text{in} \,\, \overline{\Omega}^c,\notag\\
\left[\phi_2\right] & \, = 0 & \text{on} \,\, \partial \Omega,\\
\left[\frac{\partial \phi_2}{\partial \boldsymbol{n}}\right] & \, = - \boldsymbol{m} \cdot \boldsymbol{n} + \frac{\partial \phi_1^{int}}{\partial \boldsymbol{n}} & \text{on} \,\, \partial \Omega, \notag\\
\phi_2^{ext}(\boldsymbol{x}) & \, = \mathcal{O}(\tfrac{1}{\left\| \boldsymbol{x} \right\|}) \quad \text{as} \,\, \left\| \boldsymbol{x} \right\| \rightarrow \infty,&\notag
\end{align}
with solution given by the \textit{single layer potential}
\begin{align}\label{slayer}
\phi_2 (\boldsymbol{x}) = \int_{\partial{\Omega}} g(\boldsymbol{y}) \, \mathcal{N}(\boldsymbol{x} - \boldsymbol{y}) \, \text{d}\sigma(\boldsymbol{y}),
\end{align}
with the \textit{Newtonian potential} $\mathcal{N}(\boldsymbol{x}) = \frac{1}{4\pi \left\|\boldsymbol{x} \right\|}$ and $g(\boldsymbol{y}) = \boldsymbol{m} \cdot \boldsymbol{n} - \frac{\partial \phi_1^{int}}{\partial \boldsymbol{n}}$.\\[0.1cm]
The advantage of this ansatz is twofold. Eqn.~\eqref{GR1} is a \textit{Poisson equation} with \textit{Dirichlet data} and, therefore, its \textit{Galerkin system} after FE discretization is symmetric, positive definite and sparse, and only has to be solved for \textit{free nodes}, i.e. \textit{non-boundary nodes}, see Sec.\ref{dirichlet}.\\
As pointed out in \cite{garcia2006adaptive}, the single layer potential in eqn.~\eqref{slayer} is continuous towards the boundary and less singular than the \textit{double layer potential} which arises in the \textit{ansatz of Fredkin-Koehler} and hence can be handled numerically more easily, also see Sec.~\ref{near_field}.\\
The potential \eqref{slayer} might be evaluated at boundary nodes, which provides the Dirichlet data for the Laplace equation in \eqref{GR2}. Hence, an approximation of the solution $\phi_2^{int}$ to \eqref{GR2} can be determined by evaluation of the single layer potential at boundary nodes and subsequently solving a Dirichlet problem $-\Delta \phi_2^{int} = 0$. In this connection, direct evaluation of the single layer potential at boundary nodes scales quadratically in the number of boundary nodes.\\
Our intention, however, is to evaluate \eqref{slayer} on all nodes of a tetrahedral finite element (FE) mesh within a $P1$ finite element method by a \textit{non-uniform Fourier} approach, which yields the complexity $\mathcal{O}(M + N)$, i.e. linear in the number of boundary elements and nodes of the mesh, respectively.\\
Without any restrictions, our fast evaluation scheme could also be applied for the above mentioned calculation of the Dirichlet data for \eqref{GR2}, followed by solving the arising Dirichlet Galerkin system to obtain an approximation of $\phi_2$ at the free nodes.\\
We further stress, that our approach can also be adapted for the ansatz of Fredkin and Koehler, which, however, will not be further discussed in this work.

\subsection{FEM for the Dirichlet problem}\label{dirichlet}  
For the sake of completeness, we briefly describe here the FEM for Dirichlet problems like \eqref{GR1}.\\
The variational formulation of \eqref{GR1} reads:\\
Find the potential $\phi_1$ in the Sobolev space with zero-boundary conditions, i.e. $\phi_1 \in H_0^1(\Omega)$, such that 
\begin{align}\label{varu1}
\int_{\Omega} \nabla \phi_1 \cdot \nabla v = \int_{\Omega} \boldsymbol{m} \cdot \nabla v \quad \text{for all} \,\, v \in H_0^1(\Omega). 
\end{align}
We discretize \eqref{varu1} on a tetrahedral mesh $\mathcal{T}$ with elements $T_j, \, j = 1\hdots M$ and nodes $\boldsymbol{x}_i,\, i=1\hdots N$ and use affine basis functions $\varphi_i^{(T_j)}, \, i=1 \hdots 4$ in each tetrahedron. The usual assembly process by local stiffness matrices and load vectors leads to a linear system of size $N \times N$, i.e. $\boldsymbol{S} \boldsymbol{x} = \boldsymbol{b}$. The \textit{stiffness matrix} $\boldsymbol{S}$ then has the entries  $ a_{km} =  \sum_{j=1}^M \int_{T_j} \nabla \eta_m \cdot \nabla \eta_k$, where $\eta_k,\, k=1 \hdots N$ is the \textit{nodal basis} (hat functions) of the space of $\mathcal{T}$-piecewise affine, globally continuous functions (a $N$-dimensional subspace of the Sobolev space $H^1(\Omega)$). Similar, the load vector has the entries $b_k = \sum_{j=1}^M \int_{T_j} \boldsymbol{m} \cdot \nabla \eta_k$, where $\boldsymbol{m}$ itself is assumed to be a $\mathcal{T}$-piecewise affine nodal interpolation.\\ Note that due to the known values of the solution at the boundary nodes, in our case of $\phi_1$ already equal zero, every Dirichlet system can always be rewritten to a system with homogeneous boundary conditions.  The nodal basis functions corresponding to free nodes (non-boundary nodes) form a basis of the space of $\mathcal{T}$-piecewise affine, globally continuous functions that are zero at the boundary [a finite dimensional subspace of the Sobolev space $H_0^1(\Omega)$]. Hence, we only have to solve a \textit{subsystem}, i.e.  
\begin{align}\label{poisson}
\boldsymbol{S}(fn,fn) \boldsymbol{x}(fn) = \boldsymbol{b}(fn) - (\boldsymbol{S} \boldsymbol{x}_{bn})(fn) =: \widetilde{\boldsymbol{b}}(fn),
\end{align}  
where \textit{fn} and \textit{bn} denote the $N_f$ and $N_b$ indices of \textit{free nodes} and \textit{boundary nodes}, respectively. The vector $\boldsymbol{x}_{bn}$ is understood as the vector of Dirichlet data (in the case here discussed equal to zero, thus $\widetilde{\boldsymbol{b}}(fn) = \boldsymbol{b}(fn)$ ) extended to length $N$ by zero-padding for indices of free nodes.\\
For an easily readable Matlab implementation in the $2$-dimensional case we refer to \cite{funken2011efficient}.\\
The resulting system is reduced to the size $N_f \times N_f$ and is symmetric, positive definite and sparse. The solution gives the weights of the nodal basis functions at free nodes. In our numerical tests we solve it by using an ILU-preconditioned conjugate gradient (CG) method, but \textit{algebraic multigrid preconditioned CG} or (onetime) \textit{LU decomposition} with backward substitution and exploiting the sparsity, could be used, which makes the complexity for \eqref{GR1} linear in $N_f$.    
\subsection{The single layer potential}
While $\phi_1$ is determined in linear time by an ordinary FEM for Dirichlet problems, the direct evaluation of the single-layer potential, i.e.
\begin{align}\label{slayer2}
\phi_2 (\boldsymbol{x}) = \int_{\partial{\Omega}} g(\boldsymbol{y}) \, \mathcal{N}(\boldsymbol{x} - \boldsymbol{y}) \, \text{d}\sigma(\boldsymbol{y}),
\end{align}
at boundary nodes or all nodes of a FE mesh would cost $\mathcal{O}(N_b^2)$ or $\mathcal{O}(N_b N)$ respectively, where $N_b$ is the number of nodes on the boundary and $N$ the total number of nodes in the discretized domain $\Omega$.\\
In the following we will introduce an efficient evaluation technique of \eqref{slayer} based on Fourier approximation of the Newton kernel on an auxiliary tensor grid.\\
Before we go into detail, we briefly state the main idea.\\
Note that in our $P1$ FE ansatz the first term of the function $g = \boldsymbol{m} \cdot \boldsymbol{n} - \frac{\partial \phi_1^{int}}{\partial \boldsymbol{n}}$ is piecewise affine, where the second one is constant for each surface triangle. For $\boldsymbol{m} \cdot \boldsymbol{n}$ we take the $L^2$-orthogonal projection onto the space of element-wise constant functions by taking the integral mean over surface triangles, i.e. 
$1/|S_j| \int_{S_j} \boldsymbol{m} \cdot \boldsymbol{n}$. Thus eqn. \eqref{slayer2} in its discretized form reads 
\begin{align}\label{slayer3}
\phi_2(\boldsymbol{x}_i) \approx \sum_{j = 1}^M  g_{j}\,\int_{S_j} \,\mathcal{N}( \boldsymbol{x}_i - \boldsymbol{y} )\,\text{d}\sigma(\boldsymbol{y}),\quad i=1\hdots N,
\end{align}
where the $S_j$ denote the $M$ surface triangles.\\
Following the idea in \cite{potts2003fast}, we split the kernel $N(x) := \mathcal{N}(\left\| \boldsymbol{x} \right\|) = 1/x, \, x := \left\| \boldsymbol{x} \right\|$ in a \textit{singular} and \textit{smooth part} respectively, i.e.
\begin{align}\label{splitting1}
N(x) =\big( \underbrace{N(x) - N_s(x) }_{ =:\, N_{\text{NF}} } \big) + N_s(x),
\end{align}  
where $N_s(.)$ is some approximation of $N(.)$ on an interval $[\epsilon, \beta],\, \beta > \epsilon > 0$ (see Sec.~\ref{kernel_appr}), which is defined on the whole real axis and  entirely smooth. $N_{NF}(.)$, on the other hand, is a \textit{'near field' correction}. We denote the corresponding multivariate functions by $\mathcal{N}_{s}(.) := N_{s}(\left\| . \right\| ) $ and $\mathcal{N}_{NF}(.) := N_{NF}(\left\| . \right\| ) $, respectively.\\
Our approximation scheme \eqref{slayer3} gets the form
\begin{align}\label{approx2}
\phi_2(\boldsymbol{x}_i) \approx \sum_{j = 1}^M g_{j} \int_{S_j} \, \mathcal{N}_{\text{NF}}( \boldsymbol{x}_i - \boldsymbol{y} ) \,\text{d}\sigma(\boldsymbol{y}) + \sum_{j = 1}^M g_j \int_{S_j} \, \mathcal{N}_s( \boldsymbol{x}_i - \boldsymbol{y} ) \,\text{d}\sigma(\boldsymbol{y}) =: \phi_{2}^{\text{NF}}(\boldsymbol{x}_i) + \phi_2^s(\boldsymbol{x}_i).
\end{align}
The near field part $\phi_2^{\text{NF}}$ only has to be computed for elements that have less or equal distance than $\epsilon$ to the target point $\boldsymbol{x}_i$, i.e. $\mathcal{N}_{\text{NF}}$ has small support. For the (weakly) singular cases, i.e. $\boldsymbol{x}_i \in S_j$, we will use a simple integral transformation, see Sec.~\ref{near_field}.\\
The fast computation of the part $\phi_2^s$ is achieved by approximation of the smooth kernel $\mathcal{N}_s$ by a Fourier series:\\
For the sake of simpler notation, we assume a scaled domain, i.e. $\Omega \subset (-1/4, 1/4)^3$, such that the arguments of $\mathcal{N}$ lie in $\mathbb{T} := \{\boldsymbol{x} \in \mathbb{R}^3 \mid -1/2 \leq_c \boldsymbol{x} <_c 1/2\}$.\\   
We approximate the smooth kernel $\mathcal{N}_s$ by its \textit{Fourier series} on $\mathbb{T}$, where $\leq_c$ means component-wise $\leq$, i.e.
\begin{align}\label{Fapprox}
\mathcal{N}_s(\boldsymbol{x}) \approx \mathcal{F}\mathcal{N}_s := \sum_{\boldsymbol{l} \in I_{\boldsymbol{n}}} c_{\boldsymbol{l}}(\mathcal{N}_s)\, e^{2\pi \mathrm{i} \boldsymbol{x} \cdot \boldsymbol{l}},
\end{align}  
where $I_{\boldsymbol{n}} := \{\boldsymbol{l} \in \mathbb{Z}^3\, \mid \, -\boldsymbol{n}/2 \leq_c \boldsymbol{l} \leq_c \boldsymbol{n}/2 - 1 \}$ and the \textit{Fourier coefficients} 
\begin{align}\label{Fcoeff}
c_{\boldsymbol{l}}(\mathcal{N}_s) = \int_{\mathbb{T}} \mathcal{N}_s(\boldsymbol{x}) \, e^{-2\pi \mathrm{i} \boldsymbol{x} \cdot \boldsymbol{l}}\, \text{d}\boldsymbol{x}. 
\end{align}

Inserting \eqref{Fapprox} into $\phi_2^s$ in \eqref{approx2} and exchanging summation order yields
\begin{align}\label{smooth_phi}
\phi_2^s(\boldsymbol{x}_i) = \sum_{\boldsymbol{l} \in I_{\boldsymbol{n}}} c_{\boldsymbol{l}}(\mathcal{N}_s)\, \Big( \underbrace{\sum_{j=1}^M g_{j}\,\int_{S_j} \, e^{-2\pi \mathrm{i}  \boldsymbol{y} \cdot \boldsymbol{l}} \, \text{d}\sigma(\boldsymbol{y}) }_{ =: b_{\boldsymbol{l}}}\Big)\, e^{2\pi \mathrm{i} \boldsymbol{x}_i \cdot \boldsymbol{l}} = \sum_{\boldsymbol{l} \in I_{\boldsymbol{n}}} \underbrace{d_{\boldsymbol{l}}}_{:= c_{\boldsymbol{l}}(\mathcal{N}_s)\,b_{\boldsymbol{l}}}\, e^{2\pi \mathrm{i} \boldsymbol{x}_i \cdot \boldsymbol{l}}.
\end{align}

The latter sum is a \textit{non-uniform discrete Fourier transform} (NDFT), which can be computed efficiently using FFT in $\mathcal{O}(|I_{\boldsymbol{n}}|\log |I_{\boldsymbol{n}}| + N)$ operations by so-called \textit{non-uniform fast Fourier transform} (NFFT), \cite{keiner2009using}.\\ 
The efficient computation of the tensor $\mathcal{B} = (\boldsymbol{b}_{\boldsymbol{l}})_{\boldsymbol{l} \in I_{\boldsymbol{n}}}$ will be discussed in the next section.\\[0.1cm]

Overall, the approximation scheme for $\phi_2^s$ has a similar form as the well known \textit{convolution theorem} for equispaced data, i.e.
\begin{align}\label{convtheo}
\phi_2^s = \text{NFFT}\Big( \big(c_{\boldsymbol{l}}(\mathcal{N}_s)\big)_{\boldsymbol{l} \in I_{\boldsymbol{n}}} \odot \mathcal{B} \Big),
\end{align}

where $\odot$ denotes element-wise multiplication and $\mathcal{B}$ is some generalization of an \textit{adjoint non-uniform discrete Fourier transform} \cite{potts2001fast} to an 'integrated Fourier basis', i.e.
$\int_{S_j} \, e^{-2\pi \mathrm{i}  \boldsymbol{y} \cdot \boldsymbol{l}} \, \text{d}\sigma(\boldsymbol{y})$.\\[0.1cm]

We started from the splitting $\mathcal{N} = \mathcal{N}_{NF} + \mathcal{N}_s$, where, due 
to the Fourier series approximation of $\mathcal{N}_s$, i.e. $\mathcal{F}\mathcal{N}_s$, the splitting of $\mathcal{N}$ reads now
\begin{align}\label{splitting2}
\mathcal{N} = (\mathcal{N} - \mathcal{N}_s) + \mathcal{F}\mathcal{N}_s + (\mathcal{N}_s - \mathcal{F}\mathcal{N}_s).
\end{align}  
We only take the approximation $\mathcal{N} \approx (\mathcal{N} - \mathcal{N}_s) + \mathcal{F}\mathcal{N}_s = \mathcal{N}_{NF} + \mathcal{F}\mathcal{N}_s$, introducing the error $\mathcal{N}_s - \mathcal{F}\mathcal{N}_s$, which, however, can be controlled by the size of the tensor grid, i.e. $\boldsymbol{n} = (n_1,n_2,n_3)$, and the near field $\epsilon$, cf. definition of $\mathcal{N}_{\text{NF}}$ in \eqref{splitting1}. Nevertheless, analysis of the error in connection with our choice for approximating $\mathcal{N}$ by a smooth function $\mathcal{N}_s$ in the far field region, see Sec.~\ref{kernel_appr}, will be given elsewhere.

\section{Non-uniform FFT for the single layer potential}\label{BEM-NFFT}
For the computation of the tensor $\mathcal{B}$ with entries $b_{\boldsymbol{l}} = \sum_{j=1}^M g_{j}\,\int_{S_j} e^{-2\pi \mathrm{i} \boldsymbol{y}\cdot \boldsymbol{l}} \, \text{d}\sigma(\boldsymbol{y}) $ we go similar lines as for the efficient computation of the adjoint non-uniform discrete Fourier transform (NDFT$^T$) \cite{potts2001fast}.\\
The essential step is a \textit{gridding procedure} of the data $(g_j)_{j = 1\hdots M}$ and FFT of the resulting tensor containing the 'smeared' source strengths. Hereby, gridding is done by convoluting the data with localized functions, whereas 
this is undone in Fourier space. The result is a generalization of the discrete Fourier transform to non-equispaced data \cite{greengard_nufft}.\\
First we introduce a well-localized univariate \textit{window function} $\upsilon$, e.g. a Gaussian function or Kaiser-Bessel function, see Sec.~\ref{window_fct} for more details, with a uniformly convergent Fourier series of its $1-$periodic extension, i.e.
\begin{align}\label{wfct1}
\widetilde{\upsilon}(x) := \sum_{r \in \mathbb{Z}} \upsilon(x + r).
\end{align}
For $3$ dimensions we simply take the \textit{tensor product} of the univariate functions to obtain a multivariate window function, i.e. 
\begin{align}\label{wfct2}
\widetilde{\Upsilon}(\boldsymbol{x}) := \prod_{q = 1}^3 \widetilde{\upsilon}(x^{(q)}).
\end{align}
For ease of computation, we further introduce the truncated version of $\widetilde{\Upsilon}$ with some \textit{cut-off parameter} $m \ll \min_{q = 1 \hdots 3} n_q,\,m \in \mathbb{N}, \boldsymbol{n} = (n_1,n_2,n_3)$, i.e.
\begin{align}\label{wfct3}
\widetilde{\Psi}(\boldsymbol{x}) := \prod_{q = 1}^3 \widetilde{\upsilon}(x^{(q)}) \, \chi_{[-\frac{m}{n_q},\frac{m}{n_q}]}(x^{(q)}),
\end{align}
where $\chi$ is the indicator function \footnote{$\chi_{[a,b]}(x) = 1$ for $x$ in $[a,b]$ and $0$ else.}.\\
We then compute an auxiliary tensor $\mathcal{A} = (a_{\boldsymbol{r}})_{\boldsymbol{r} \in I_{\alpha\,\boldsymbol{n}}}$, where $\alpha > 1$ is an \textit{over-sampling factor}, i.e.
\begin{align}\label{gridding}
a_{\boldsymbol{r}} := \sum_{j=1}^M g_{j}\,\int_{S_j} \, \widetilde{\Psi}(\boldsymbol{r} \odot (\alpha\,\boldsymbol{n})^{-1} - \boldsymbol{y}) \, \text{d}\sigma(\boldsymbol{y}),
\end{align}
where $\odot$ denotes element-wise multiplication and $(\alpha\,\boldsymbol{n})^{-1}$ is meant component-wise and corresponds to the mesh size of the auxiliary tensor grid, see Fig.~\ref{indx}.\\
Formula \eqref{gridding} can be seen as \textit{gridding} of the source strengths $g_j$ on an auxiliary tensor grid of size $|I_{\alpha \,\boldsymbol{n}}|$. The desired tensor $\mathcal{B} = (b_{\boldsymbol{l}})_{\boldsymbol{l} \in I_{\boldsymbol{n}}}$ in \eqref{smooth_phi} can be computed by the Fourier transform of $\mathcal{A}$. More precisely, we define a function $f$ according to the definition of $\mathcal{A}$ by
\begin{align}\label{fx}
f(\boldsymbol{x}) := \sum_{j=1}^M \,g_{j}\,\int_{S_j} \, \widetilde{\Upsilon}(\boldsymbol{x} - \boldsymbol{y}) \, \text{d}\sigma(\boldsymbol{y}).
\end{align}
By expressing the Fourier coefficients of $f$ in two different ways, we will end up with a simple formula for computing the tensor $\mathcal{B}$.\\
First we approximate the Fourier coefficients of $f$ (cf.~\eqref{fx}) by the trapezoidal rule for $\boldsymbol{l} \in I_{\boldsymbol{n}}$ and $\widetilde{\Upsilon}$ by the truncated version $\widetilde{\Psi}$ in \eqref{wfct3}, i.e.
\begin{align}\label{FFT}
c_{\boldsymbol{l}}(f) = \int_{\mathbb{T}} f(\boldsymbol{x})\, e^{-2\pi \mathrm{i} \boldsymbol{x} \cdot \boldsymbol{l}} \, \text{d}\boldsymbol{x}
\approx \frac{1}{|I_{\alpha\,\boldsymbol{n}}|} \, \sum_{\boldsymbol{r} \in I_{\alpha\,\boldsymbol{n}}} \Big( \underbrace{\sum_{j=1}^M  \,g_{j}\,\int_{S_j} \, \widetilde{\Psi}(\boldsymbol{r} \odot (\alpha\,\boldsymbol{n})^{-1} - \boldsymbol{y})\, \text{d}\sigma(\boldsymbol{y}) }_{ = a_{\boldsymbol{r}}} \Big)\, e^{-2\pi \mathrm{i}  (\boldsymbol{r} \odot (\alpha\,\boldsymbol{n})^{-1}) \cdot\boldsymbol{l}},
\end{align}
which can be computed by a multivariate FFT of the tensor $\mathcal{A}$.\\

On the other hand, we also obtain an approximation of $c_{\boldsymbol{l}}(f)$ by inserting the truncated Fourier series of $\widetilde{\Upsilon}$, i.e.
\begin{align}
\widetilde{\Upsilon}(\boldsymbol{x}) \approx \sum_{\boldsymbol{l} \in I_{\boldsymbol{n}}} c_{\boldsymbol{l}}(\widetilde{\Upsilon}) \, e^{2\pi \mathrm{i} \boldsymbol{x} \cdot \boldsymbol{l}},
\end{align}
into the expression for the function $f$, i.e.
\begin{align}
f(\boldsymbol{x}) & \,\approx \sum_{\boldsymbol{l} \in I_{\boldsymbol{n}}} \,  \Big( \underbrace{\sum_{j=1}^M g_{j}\, c_{\boldsymbol{l}}(\widetilde{\Upsilon}) \, \int_{S_j} \,e^{-2\pi \mathrm{i} \boldsymbol{y} \cdot \boldsymbol{l}} \, \text{d}\sigma(\boldsymbol{y})}_{ = c_{\boldsymbol{l}}(f)} \Big) \, e^{2\pi \mathrm{i} \boldsymbol{x} \cdot \boldsymbol{l}} \\[0.1cm] 
& \, = \sum_{\boldsymbol{l} \in I_{\boldsymbol{n}}} \,\Big( c_{\boldsymbol{l}}(\widetilde{\Upsilon}) \, \underbrace{\sum_{j=1}^M  g_{j}\,\int_{S_j} \,e^{-2\pi \mathrm{i} \boldsymbol{y} \cdot \boldsymbol{l}} \, \text{d}\sigma(\boldsymbol{y})}_{ = b_{\boldsymbol{l}}} \Big)\, e^{2\pi \mathrm{i} \boldsymbol{x} \cdot \boldsymbol{l}}.
\end{align}
Thus, we have the relation ($\boldsymbol{l} \in I_{\boldsymbol{n}}$)
\begin{align}\label{deconv}
b_{\boldsymbol{l}} =  c_{\boldsymbol{l}}(f)/c_{\boldsymbol{l}}(\widetilde{\Upsilon}).
\end{align}
Overall the computation of $\mathcal{B}$ consists of computing the coefficients $c_{\boldsymbol{l}}(f)$ in \eqref{FFT} by a multivariate FFT of the gridding tensor $\mathcal{A}$, followed by element-wise division by the precomputed coefficients $c_{\boldsymbol{l}}(\widetilde{\Upsilon})$.\\
We therefore conclude that these two steps together scale $\mathcal{O}(|I_{\alpha \boldsymbol{n}}| \, \log(|I_{\alpha \boldsymbol{n}}|) + |I_{\boldsymbol{n}}|)$. As will be shown in the next section, the computation of the gridding tensor $\mathcal{A}$
can be done linearly in the number of surface elements, i.e. $\mathcal{O}(M)$. Hence, in total, computing $\mathcal{B}$ scales $\mathcal{O}\big( M + |I_{\alpha \boldsymbol{n}}| \, \log(|I_{\alpha \boldsymbol{n}}|) \big)$.\\ 
We stress that, alternatively to the above procedure for computing the tensor $\mathcal{B}$, we also could have directly transformed the expression into a discrete sum by using quadrature, i.e
\begin{align}\label{naive_gridding}
b_{\boldsymbol{l}} = \sum_{j=1}^M g_{j}\,\int_{S_j} e^{-2\pi \mathrm{i} \boldsymbol{y}\cdot \boldsymbol{l}} \, \text{d}\sigma(\boldsymbol{y}) \approx \sum_{j=1}^M \sum_{s = 1}^{Q_j} \omega_{j,s} g_{j}\,e^{-2\pi \mathrm{i} \boldsymbol{y}_{j,s} \cdot \boldsymbol{l}} \equiv \sum_{k = 1}^Q \widetilde{\omega}_{k}\,e^{-2\pi \mathrm{i} \boldsymbol{y}_{k} \cdot \boldsymbol{l}},  
\end{align}
where $k$ is a long index, e.g. $k = j + (s-1)M$. Eqn.~\eqref{naive_gridding} could then be computed by a standard adjoint NFFT \cite{keiner2009using} in $\mathcal{O}\big(Q + |I_{\alpha \boldsymbol{n}}| \, \log(|I_{\alpha \boldsymbol{n}}|)\big),\, Q:= \sum_{j=1}^M Q_j$ operations, also compare with \cite{kritsikis2008fast}. Since the number of quadrature points $Q$ might be very large, this approach is rather impractical.   
For that reason, we use eqn.~\eqref{deconv} for the computation of the coefficients $b_{\boldsymbol{l}}$, where we can precompute the integrals in a setup phase of a micromagnetic simulation, compare with Alg.~\ref{alg_tot}.\\ 
However, at least it gives us a direct analogy to the standard adjoint NFFT. In particular, the choice of window functions can be justified, since basically the same error estimates with respect to the cut-off parameter $m$ and over-sampling factor $\alpha$ hold for our method, see Sec.~\ref{window_fct}. 
\subsection{Computation of the gridding tensor}\label{Gridding}
We take a closer look at the computation of the tensor $\mathcal{A}$ (compare with \eqref{gridding}), i.e.
\begin{align}\label{gridding2}
a_{\boldsymbol{r}} = \sum_{j=1}^M g_{j}\,\int_{S_j} \, \widetilde{\Psi}(\boldsymbol{r} \odot (\alpha\,\boldsymbol{n})^{-1} - \boldsymbol{y}) \, \text{d}\sigma(\boldsymbol{y}).
\end{align}
The aim is to compute \eqref{gridding2} through sparse summation by exploiting the locality of the function $\widetilde{\Psi}$.\\
We further assume the domain scaled into the hypercube $(-1/4,1/4)^3$, hence we also have $\Omega \subset \mathbb{T}$.\\ 
A triangle of the surface mesh is given as $S_j \equiv \{\boldsymbol{y}_{0,j},\hdots,\boldsymbol{y}_{2,j},\, \boldsymbol{y}_{k,j} \neq \boldsymbol{y}_{l,j}, \,\,\text{for}\,\,k\neq l\}$ where
\begin{align}\label{triangle}
\boldsymbol{y} \in S_j \Leftrightarrow \exists \, \xi_1,\xi_2 \in \Delta_0: \boldsymbol{y} = \boldsymbol{y}_{0,j} + \xi_1 (\boldsymbol{y}_{1,j}-\boldsymbol{y}_{0,j}) + \xi_2 (\boldsymbol{y}_{2,j}-\boldsymbol{y}_{0,j}),
\end{align}
where $\Delta_0$ denotes the unit triangle in $2$d.\\
In order to achieve linear complexity in $M$ we define a subset of $I_{\alpha\boldsymbol{n}}$ for each surface element $S_j$ that ensures that $\boldsymbol{r} \odot (\alpha\,\boldsymbol{n})^{-1} - \boldsymbol{y}$ in \eqref{gridding2} lies in the hypercube $\varprod_{q=1}^3 [-m n_q^{-1},m n_q^{-1}]$, i.e.
\begin{align}\label{indexset}
I_{\alpha\,\boldsymbol{n},m}(S_j) & \,:=  \{\boldsymbol{l} \in I_{\alpha\,\boldsymbol{n}} \mid -m\boldsymbol{n}^{-1} \leq_c (\alpha\,\boldsymbol{n})^{-1} \odot \boldsymbol{l} - \boldsymbol{y} \leq_c m\boldsymbol{n}^{-1},\, \boldsymbol{y} \in S_j \} \\[0.1cm]
& \, = \{\boldsymbol{l} \in I_{\alpha\,\boldsymbol{n}} \mid \boldsymbol{y} \odot \alpha\,\boldsymbol{n} - m\boldsymbol{1} \leq_c \boldsymbol{l} \leq_c \boldsymbol{y} \odot \alpha\,\boldsymbol{n} + m\boldsymbol{1},\, \boldsymbol{y} \in S_j \}. 
\end{align}

\begin{figure}[hbtp]
\center
\includegraphics[scale = 0.8]{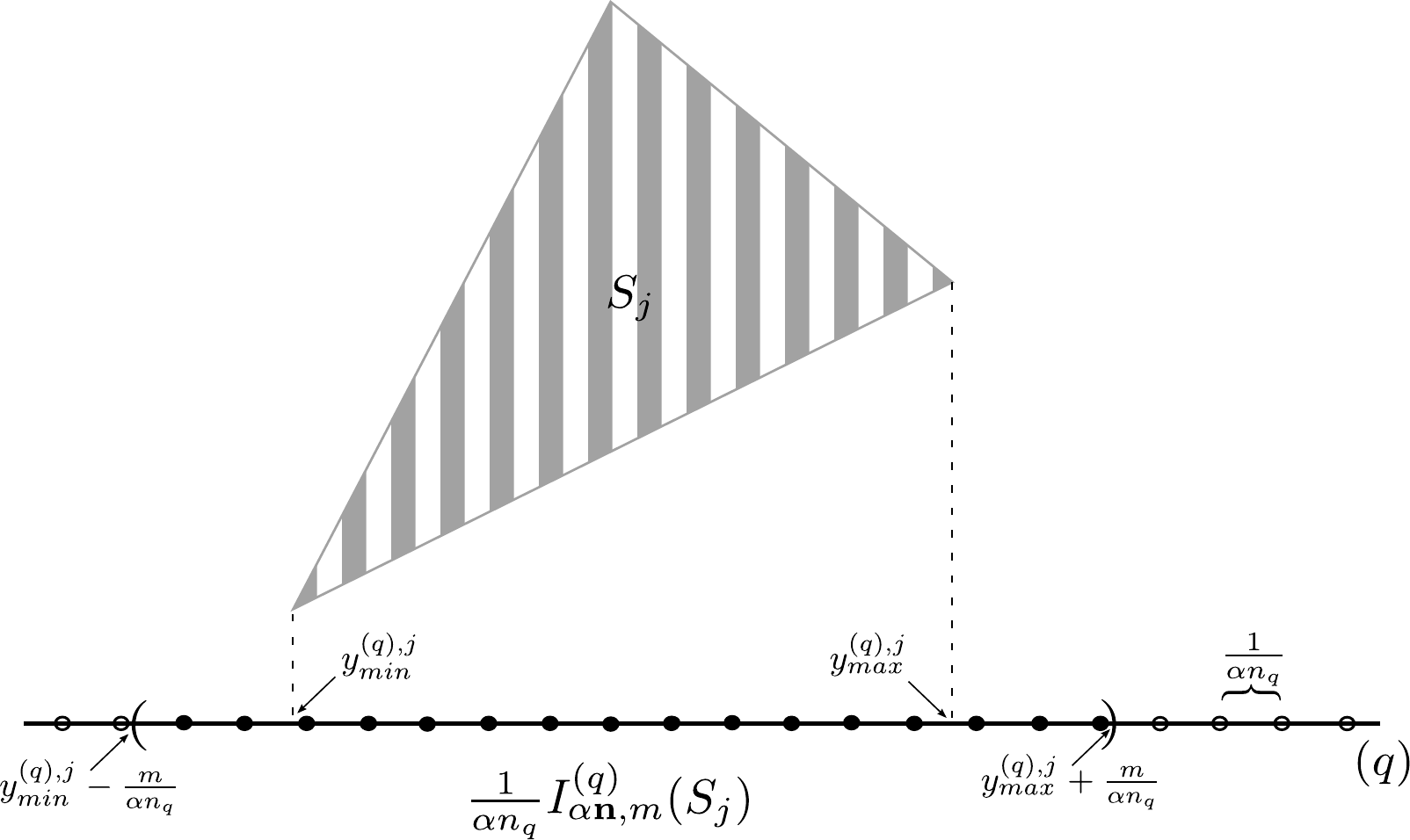}    
\caption{The $q$-th component of the index set $I_{\alpha\boldsymbol{n},m}(S_j)$ (filled dots). $q$ denotes the space direction, $S_j$ is one surface triangle, $\boldsymbol{n} = (n_1,n_2,n_3)$ the size of the tensor grid, $\alpha$ the over-sampling factor, $m$ the cut-off parameter.
$y_{\text{min}/\text{max}}^{(q),j}$ is the left and right most corner of the triangle, respectively. $1/(\alpha n_q)$ is the mesh size of the grid in the $q$-th direction.}\label{indx}
\end{figure}

We denote the $q-$th component of \eqref{indexset} by $I_{\alpha\,\boldsymbol{n},m}^{(q)}(S_j)$, where we take the $q-$th components of the vector expressions in the definition.\\
For the sake of computation we rewrite 
\begin{align}\label{indexset2}
I_{\alpha\boldsymbol{n},m}^{(q)}(S_j) = \{l_q \in I_{\alpha\,\boldsymbol{n}}^{(q)} \mid \alpha \,n_q\,\min_{\boldsymbol{y} \in S_j} y^{(q)}  - m \leq l_q \leq  \alpha\,n_q\,\max_{\boldsymbol{y} \in S_j} y^{(q)} + m\}. 
\end{align}
From \eqref{triangle} it is easily seen that for the expressions $y_{\text{min}}^{(q),j} := \min_{\boldsymbol{y} \in S_j} y^{(q)}$ and $y_{\text{max}}^{(q),j} := \max_{\boldsymbol{y} \in S_j} y^{(q)}$ in \eqref{indexset2} simply holds
\begin{align}\label{index_opt_sol}
y_{\text{min}}^{(q),j} & \, = \min_{k=0,1,2 } y_k^{(q),j} \\[0.1cm]
y_{\text{max}}^{(q),j} & \, = \max_{k=0,1,2} y_k^{(q),j}. 
\end{align}
Due to our assumption $\Omega \subset \mathbb{T}$, we have $|I_{\alpha\,\boldsymbol{n},m}^{(q)}(S_j)| \leq 2m + 1 + \alpha\,n_q\,\max_{j=1\hdots M}|y_{\text{max}}^{(q),j} - y_{\text{min}}^{(q),j}| =: \widetilde{m}_q$ and $|I_{\alpha\,\boldsymbol{n},m}(S_j)| \leq \prod_{q=1}^3 \widetilde{m}_q =: \mu$.
Fig.~\ref{indx} shows the index set $I_{\alpha\boldsymbol{n},m}^{(q)}(S_j)$.\\[0.1cm]
The tensor $\mathcal{A}$ in \eqref{gridding2} is now computed by only using the index sets $I_{\alpha\boldsymbol{n},m}(S_j)$ in $\mathcal{O}(\mu M)$ operations:
\begin{itemize}
\item Initialize $\mathcal{A}$ with zeros
\item For $j = 1 \hdots M$ calculate the vector $\Big(g_{j}\,\int_{S_j} \, \widetilde{\Psi}(\boldsymbol{l} \odot \boldsymbol{n}^{-1} - \boldsymbol{y}) \, \text{d}\sigma(\boldsymbol{y})\Big)_{\boldsymbol{l} \in I_{\alpha\,\boldsymbol{n},m}(S_j)}$ of length at most $\mu$ and add the corresponding components to $\mathcal{A}$.
\end{itemize}
Here, the integrals are precomputed, since they only depend on the given mesh. We may store the sparse matrix
\begin{align}\label{A}
 \boldsymbol{A} := \Big(\int_{S_j} \, \widetilde{\Psi}(\boldsymbol{l} \odot \boldsymbol{n}^{-1} - \boldsymbol{y}) \, \text{d}\sigma(\boldsymbol{y})\Big)_{j = 1\hdots M,\, \boldsymbol{l} \in I_{\alpha\,\boldsymbol{n},m}(S_j)}.
\end{align}
Nevertheless, since the integrals of \eqref{A} are smooth functions in the parameter $\boldsymbol{l}$, we can think of tensor compression for the rows, i.e. $\boldsymbol{A}(j,:) \in \bigotimes_{q=1}^3 \mathbb{R}^{I_{{\alpha\,\boldsymbol{n},m}}^{(q)}(S_j)}$, 
reducing the storage to $\mu^\prime M$ for $\mu^\prime < \mu$ depending on the tensor format and the accuracy. Tab.~\ref{tab1} shows examples for compression rates using tensor train (TT) approximation \cite{tt_tensor2}. 
\begin{table}
  \centering
  \begin{tabular}{c c c c}
    $\#$ surface elements & $m$ & full (mb)  & compressed (mb) \\ \hline\hline
    $1.3$e$3$   &  $5$  &  $48$  & $14$     \\\hline
    $2.6$e$3$   &  $5$  &  $77$ & $20$     \\\hline
    $1.3$e$3$   &  $6$  &  $73$  & $15$     \\\hline
    $2.6$e$3$   &  $6$  &  $120$ & $21$     \\ \hline
 \end{tabular}
  \caption{Compression of $\boldsymbol{A}$ from surface mesh of a sphere by the Tensor Train format with accuracy $1$e-$8$ measured in the relative Frobenius norm. $n_q \equiv 72, \, \alpha = 2$.}\label{tab1} 
\end{table}
\subsection{Window functions}\label{window_fct}
\begin{figure}[hbtp]
\center
\includegraphics[scale = 0.3]{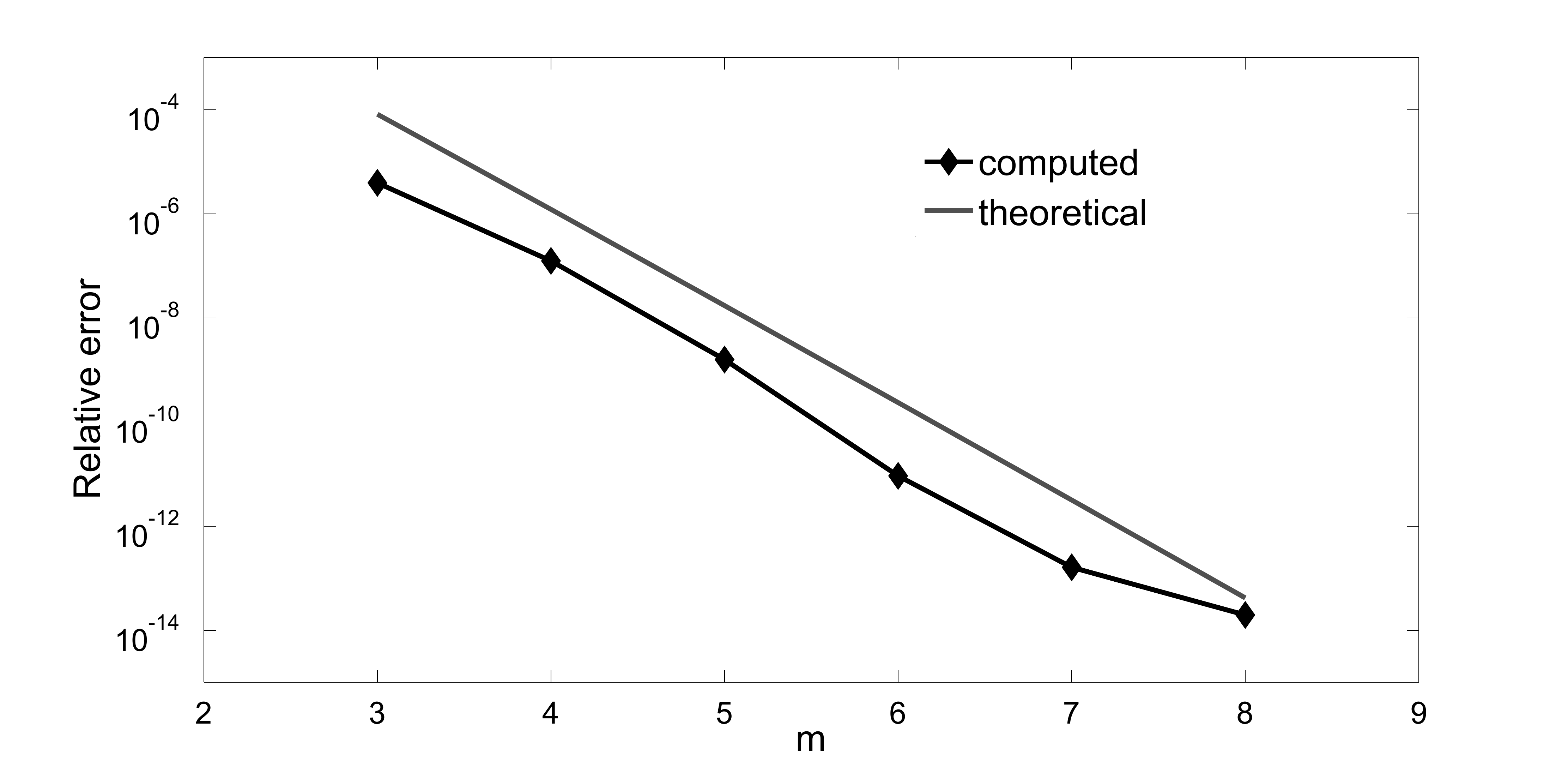}    
\caption{Relative error for the computation of $\mathcal{B}$ using Kaiser-Bessel functions, as defined in \eqref{KBfct}, on a triangular mesh ($M \sim 3e2$) of the surface of a sphere and randomly chosen values $g_j \in [-1,1]$, $\alpha = 2$ and $n_q \equiv 32, q=1\hdots 3$.}\label{gridding_m}
\end{figure}

In \cite{elbel_1998,potts_habil} it was shown that, in the case of \textit{Gaussian, Sinc, cardinal B-splines} or \textit{Kaiser-Bessel} window functions, the error for (adjoint) NFFT decays exponentially in the cut-off parameter $m$.\\
Hereby, Kaiser-Bessel functions have the fastest decaying error bound. For $n \in 2\mathbb{N}$ we define the univariate Kaiser-Bessel function 
\begin{align}\label{KBfct}
\upsilon(x) := \left\{
\begin{array}{c c}
 \frac{\sinh(b \sqrt{m^2 - (\alpha n)^2 x^2})}{\pi \sqrt{m^2 - (\alpha n)^2 x^2}}, & |x| \leq \frac{m}{\alpha n} \\[0.1cm]
 \frac{b}{\pi}, & |x| = \frac{m}{\alpha n} \\[0.1cm]
 \frac{\sin(b \sqrt{(\alpha n)^2 x^2 - m^2})}{\pi \sqrt{(\alpha n)^2 x^2-m^2}}, & \text{else},
\end{array}\right.
\end{align}
where $b := \pi(2-1/\alpha)$. The Fourier coefficients are given by
\begin{align}\label{cFB}
 c(\upsilon)(k) = \left\{
\begin{array}{c c}
 \frac{1}{\alpha n} I_0(m \sqrt{b^2 - (\frac{2\pi k}{\alpha n})^2}), & |k| \leq \alpha n (1 - \frac{1}{2\alpha})\\[0.1cm]
0, & \text{else}, \\[0.1cm]
\end{array}\right.
\end{align}
where $I_0$ is the \textit{modified zero order Bessel-function of the first kind}.\\
For the univariate setting a bound for the relative error produced by NFFT with Kaiser-Bessel functions is \cite{potts_habil}
\begin{align}\label{error_bound_nufft}
 C(\alpha,m) = 4\pi (\sqrt{m} + m) (1-1/\alpha)^{1/4} \exp(-2\pi m \sqrt{1-1/\alpha}),
\end{align}
which already indicates small errors for $m$ about $4$ and $\alpha = 2$, see Fig.~\ref{gridding_m}.\\
Note that this error bound is independent of $n$ and $M$. The error decays exponentially with increasing $m$, but not with increasing $\alpha$. Therefore, we 
fix $\alpha = 2$ throughout the paper and control the error by choosing the cut-off $m$ appropriately.\\
Since our method for computing $\mathcal{B}$ is mathematically equivalent to a NFFT if just accurate enough quadrature is used [compare with \eqref{naive_gridding}], we compare with the theoretical error bound \eqref{error_bound_nufft} for standard NFFT. In this context, also note that the computation of $\mathcal{B}$ is stable regarding round off errors \cite{potts_habil}. Fig.\ref{gridding_m} shows the cut-off parameter $m$ versus the relative error in the maximum-norm, i.e. $\max_{\boldsymbol{l}}|\mathcal{B} - \mathcal{B}_{\text{exact}}| / \max_{\boldsymbol{l}}| \mathcal{B}_{\text{exact}}|$, 
for a triangular mesh of the surface of a sphere, randomly chosen values $g_j \in [-1,1]$ and $\alpha = 2$. 

\subsection{Kernel approximation}\label{kernel_appr}
\begin{figure}[hbtp]
\center
\includegraphics[scale = 0.3]{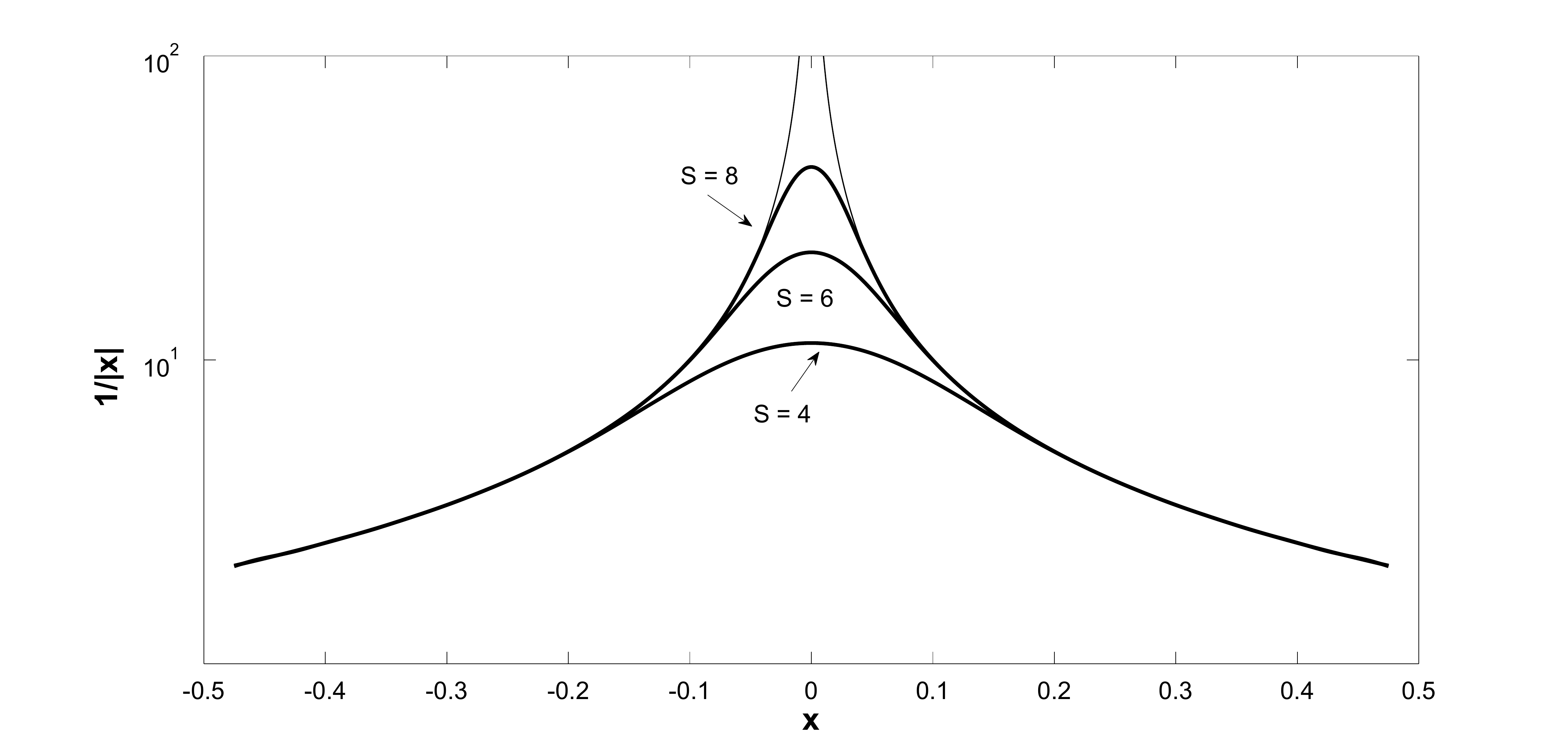}    
\caption{Approximation of $N(x) = 1/|x|$ by exponential sums.}\label{expsum_fig}
\end{figure}

We now turn to the approximation of the Newtonian kernel $\mathcal{N}$ in a region $[\epsilon, \beta],\, \beta > \epsilon > 0$, where we set $\beta = 1/2$ due to our scaling convention $\Omega \subset (-1/4,1/4)^3$.\\
As described in \cite{braess2009efficient} the kernel $N(x) = 1/|x|$ can be approximated by exponential sums in an interval $[1,R]$, i.e.
\begin{align}\label{expsum}
N(x) \approx N_s(x) := \sum_{k=1}^S \omega_k e^{-\gamma_k x^2} \equiv \sum_{k=1}^S \omega_k\, N_s^{(k)}(x),
\end{align}
where the weights $\omega_k$ and nodes $\gamma_k$ were computed for several configurations of the parameters $R, S$ and uniform absolute error bound $err$. A simple transformation of the weights and nodes yields a corresponding approximation
on the desired interval $[\epsilon,1/2]$, i.e.
\begin{align}
\omega_{\text{trans}} =& \, \omega/h_{\text{min}}\\
\gamma_{\text{trans}} =& \, \gamma/h_{\text{min}}^2\\
err_{\text{trans}} =& \, err/h_{\text{min}},
\end{align}
where $h_{\text{min}} := 1/(2\sqrt{R})$.\\
For our numerical tests we choose the computed values for $1/\sqrt{x}$ with $S=21$ and $R=7e4$ from \cite{exp_sums_hackbusch} 
, yielding a uniform error of $5.79e-06$ in $[1.89e-03,5.00e-01]$. 
Depending on the actual near field $\epsilon$ we truncate the expansion \eqref{expsum}, taking only $S^\prime \leq S$ terms, in order to have an accurate approximation only in the sub-interval $[\epsilon, 1/2]$. Fig.~\ref{expsum_fig} shows the smooth approximation $N_s(.)$ for different 
number of terms in the expansion \eqref{expsum}.

Note that $\mathcal{N}_s$ and hence its Fourier transform are both sums of separable functions, i.e. 
\begin{align}
\mathcal{N}_s(\boldsymbol{x}) = & \,\sum_{k=1}^{S^\prime} \omega_k\,N_s^{(k)}(x_1) \,N_s^{(k)}(x_2)\, N_s^{(k)}(x_3) \\
\mathcal{F}\mathcal{N}_s(\boldsymbol{r}) = & \,\sum_{k=1}^{S^\prime} \omega_k\,\mathcal{F}N_s^{(k)}(r_1) \,\mathcal{F}N_s^{(k)}(r_2)\, \mathcal{F}N_s^{(k)}(r_3).
\end{align}
This fact allows us to store only $S^\prime \sum_{q=1}^3 n_q$ complex numbers, 
instead of $\prod_{q=1}^{3} n_q$ for all Fourier coefficients of the multivariate function $\mathcal{N}_s$.
However, additional $\mathcal{O}(S^\prime)$ operations have to be performed on runtime to calculate one coefficient from its factorized representation.

\subsection{Near field correction}\label{near_field}
The near field correction is determined by (cf. \eqref{approx2})
\begin{align}\label{nfcorr}
\phi_2^{NF}(\boldsymbol{x}_i) \approx \sum_{j = 1}^M g_{j} \int_{S_j} \, \mathcal{N}_{\text{NF}}( \boldsymbol{x}_i - \boldsymbol{y} ) \,\text{d}\sigma(\boldsymbol{y}). 
\end{align}
Since $\mathcal{N}_{\text{NF}}$ has small support, \eqref{nfcorr} only has to be computed for surface elements that have less or equal distance than $\epsilon$ to the target point $\boldsymbol{x}_i$, i.e. for summation we only use the \textit{index sets} 
\begin{align}\label{nf_index}
I_{\text{NF}_{\epsilon}}(\boldsymbol{x}_i) := \{ j \in \{1, \hdots, M\} \,\mid \, \text{d}(\boldsymbol{x}_i,S_j) \leq \epsilon \}.
\end{align} 
It is easily seen that
\begin{align}
 \bigcup_{i=1}^N   I_{\text{NF}_{\epsilon}}(\boldsymbol{x}_i) \times i = \bigcup_{j=1}^M   j \times I_{\text{NF}_{\epsilon}}(S_j),  
\end{align}
where %
\begin{align}\label{nf_index2}
I_{\text{NF}_{\epsilon}}(S_j) := \{ i \in \{1, \hdots, N\} \,\mid \, \text{d}(\boldsymbol{x}_i,S_j) \leq \epsilon \}.
\end{align} 
Using this relation we can sum up \eqref{nfcorr} in a similar way like the gridding tensor in $\mathcal{O}(M)$ operations, cf. Sec.\ref{Gridding}, i.e.
\begin{itemize}
\item Initialize $\phi_2^{NF}$ with zeros
\item For $j = 1 \hdots M$ calculate the vector $\Big(g_{j}\,\int_{S_j} \, \mathcal{N}_{\text{NF}}( \boldsymbol{x}_i - \boldsymbol{y} )  \, \text{d}\sigma(\boldsymbol{y})\Big)_{i \in I_{\text{NF}_{\epsilon}}(S_j)}$ 
and add the corresponding components to $\phi_2^{NF}$.
\end{itemize}
The integrals are again precomputed and stored in a sparse matrix. For reasonably uniform distribution of nodes nearby the boundary, we may assume that $\nu := \max|I_{\text{NF}_{\epsilon}}(S_j)|$ is much smaller than $N$.
The complexity of the calculation of $\phi_2^{NF}$ is therefore at most $\mathcal{O}(\nu M)$.\\
For the (weakly) singular cases in \eqref{nfcorr}, i.e. $\boldsymbol{x}_i \in S_j$, we use the following substitutions. Assume $S_j$ has the vertices $\boldsymbol{x}_1, \boldsymbol{x}_2$ and $\boldsymbol{x}_3$
and we want to evaluate at $\boldsymbol{x}_2$. We parameterize $S_j$ by 
\begin{align}
 p:\, \Delta_0 \rightarrow \mathbb{R}^3, (s,t) \mapsto \boldsymbol{x}_2 + s(\boldsymbol{x}_1 - \boldsymbol{x}_2) + t(\boldsymbol{x}_3 - \boldsymbol{x}_1),
\end{align}
where $\Delta_0$ is the unit triangle in the plane. After the substitution $s \rightarrow s$ and $t \rightarrow st$ with Jacobian determinant $|J| = s$ the integration domain gets the unit square in the plane and the integral gets non-singular. We then treat it by tensor product Gaussian quadrature.\\
For more information we refer to \cite{lyness1994survey}.
\section{Numerics}
\begin{figure}[hbtp]
\center
\includegraphics[scale = 0.45]{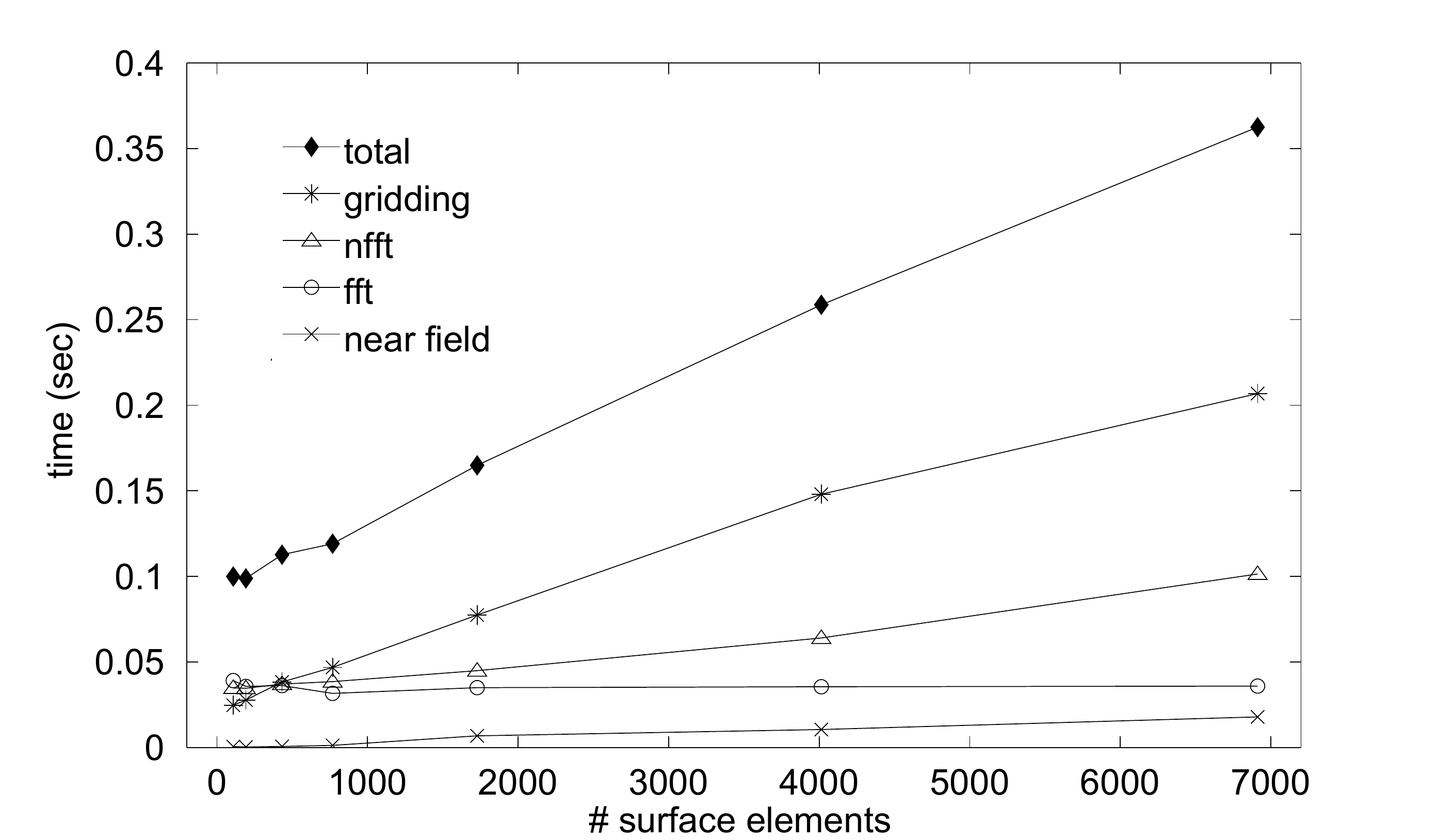}    
\caption{Cpu-times (sec) versus number of surface elements for the calculation of the single-layer potential in a cube. $n_q \equiv 48, \, \alpha = 2$ and $m = 5$.}\label{slayer_times1}
\end{figure}

\begin{figure}[hbtp]
\center
\includegraphics[scale = 0.45]{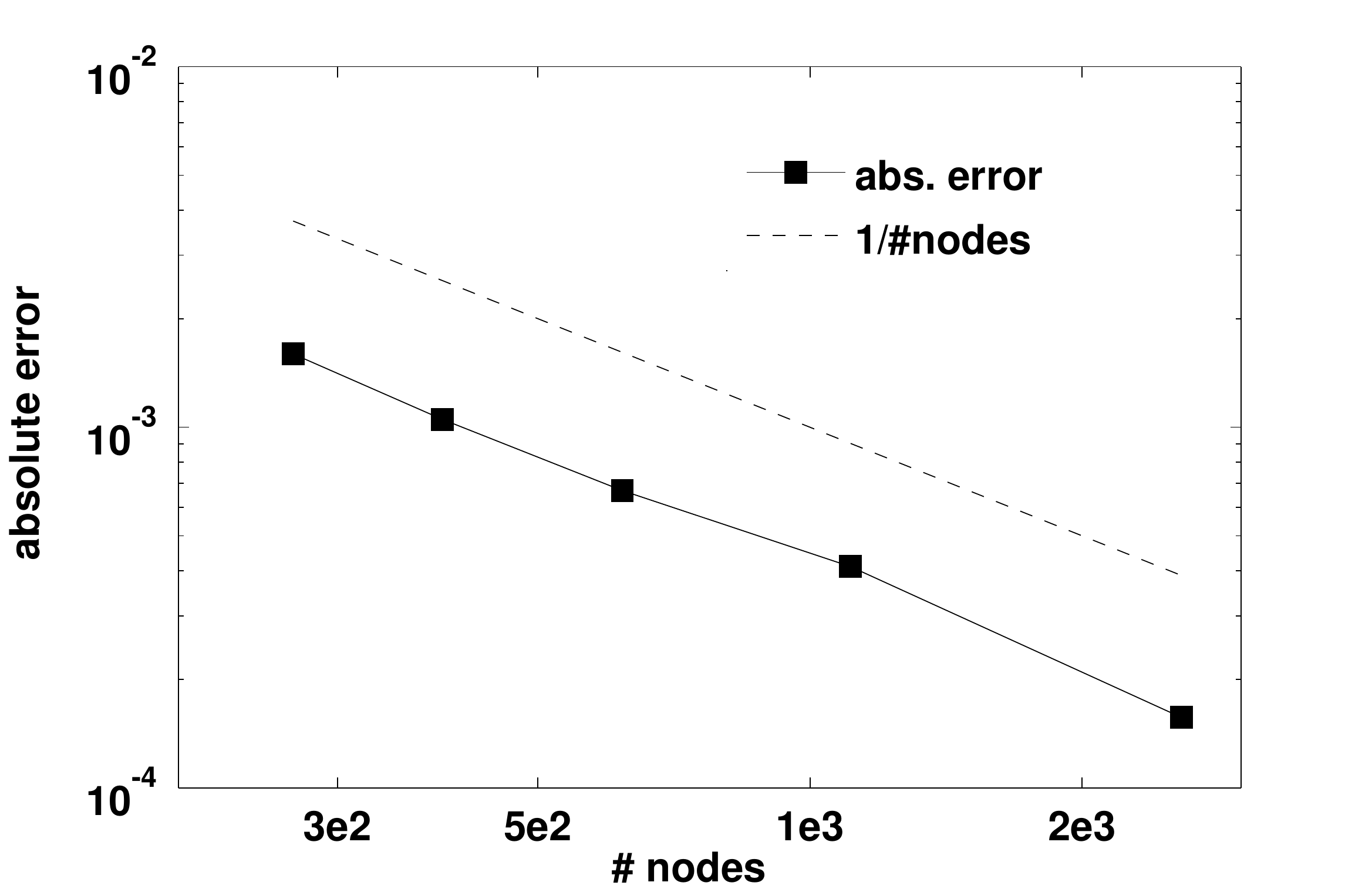}    
\caption{Maximum of the absolute error for uniform magnetization in a sphere with radius $0.2$ and center at zero.}\label{error1}
\end{figure}

The tests were taken on a Linux Workstation with a hexa-core AMD Phenom II X6 1090T processor and 16 GB RAM. We used Matlab $7.13.0$ and the C library NFFT 3 \cite{keiner2009using}.\\ 
Alg.~\ref{alg_tot} shows a pseudo-code of the described method for solving problem \eqref{scpot_classical} by the ansatz \eqref{GR1} and \eqref{GR2} using our proposed fast evaluation scheme for the single layer potential. Whereby, 
we divide the total algorithm into a setup and a computation phase. In any micromagnetic solver the computation phase is part of the effective field evaluation, which has to be done at every step of the iterative solution procedure. 
The setup phase only depends on the geometry of the problem and thus has to be done only once for a given problem. In the following, we show that the computational effort for the computation phase scales linearly with the problem size.
In Alg.~\ref{alg_tot} for computing the magnetic scalar potential the first step of the computation phase is the solution of a Dirichlet problem for $\phi_1$. Since the problem is sparse and the LU decomposition is done in a setup phase the complexity is linear.
The numerical experiments in this section also show linear complexity for the computation of $\phi_2$.\\
\begin{algorithm}
\caption{Scalar potential}
\label{alg_tot}
\begin{algorithmic}
\Require $\boldsymbol{m}\in \big( H^1 (\Omega) \big)^3$, \text{mesh}\,\, $\mathcal{T}$ of $\Omega \subset (-1/4,1/4)^3 \subset \mathbb{T}$, $\boldsymbol{n} \in 2\mathbb{N}^3$, $\epsilon>0$, $m \in \mathbb{N}$, $\alpha \in \mathbb{N}, \alpha \geq 2$ 
\Ensure $\phi^{int} \in H^1 (\Omega)$\\
\textbf{Setup}
\begin{itemize}
  \item \text{Compute the LU decomposition of the stiffness matrix and the linear operator for the RHS, cf. Sec.~\ref{dirichlet}}
  \item \text{Compute the matrix $\boldsymbol{A}$ from \eqref{A}, cf. Sec.~\ref{Gridding}}
  \item \text{Compute the Fourier coefficients of the window functions from \eqref {cFB}, cf. Sec.~\ref{window_fct}} 
  \item \text{Compute the Fourier coefficients of the Kernel approximation, i.e. $\big(c_{\boldsymbol{l}}(\mathcal{N}_s)\big)_{\boldsymbol{l} \in I_{\boldsymbol{n}}}$, cf. Sec.~\ref{kernel_appr}}
  \item \text{Compute the integrals of the near field correction, cf. Sec.~\ref{near_field}}
\end{itemize}
\textbf{Actual computation}
\begin{itemize}
\item \text{Solve the linear system \eqref{poisson} for $\phi_1^{int}$}
\item \text{Compute $\phi_2^{s}$:}
  \begin{itemize}
  \item \text{Compute the tensor $ \mathcal{A}$ in \eqref{gridding2}} 
  \item \text{Compute the multivariate FFT of the tensor $\mathcal{A}$, cf. \eqref{FFT}}
  \item \text{Compute the tensor $\mathcal{B}$ by formula \eqref{deconv}}
  \item \text{Compute $\mathcal{D} := \big(c_{\boldsymbol{l}}(\mathcal{N}_s)\big)_{\boldsymbol{l} \in I_{\boldsymbol{n}}} \odot \mathcal{B}$}
  \item \text{Compute the NFFT of $\mathcal{D}$ to obtain $\phi_2^{s}$, cf. \eqref{convtheo}}
  \end{itemize}
\item \text{Compute $\phi_2^{NF}$ as described in Sec.~\ref{near_field}}
\end{itemize}
\State $\phi_2^{int} \gets \phi_2^s + \phi_2^{NF}$
\State $\phi^{int} \gets \phi_1^{int} + \phi_2^{int}$ 
\end{algorithmic}
\end{algorithm} 
We first tested our method for computing the single layer potential for a cube. Fig. \ref{slayer_times1} shows the cpu-times in seconds of the different parts of our algorithm for randomly chosen values $g_j \in [-1,1]$. The parts \textit{gridding} and \textit{fft} correspond to the computation of the tensor $\mathcal{B}$, compare with \eqref{deconv}, where
times for the element-wise division with the precomputed Fourier coefficients of the window function were included in the times for the FFT. We do not give the times for the element-wise multiplication of the Fourier coefficients of $\mathcal{N}_s$ and $\mathcal{B}$, since they are negligible. For the NFFT we used the C library NFFT $3$. 
We set $m = 5$ and $\alpha = 2$, both, in our gridding method as well as in the NFFT and used $n_q \equiv 48$ for this tests. Further we have chosen $\epsilon$ such that $\nu$ in the complexity of the near field correction was below $3$e$2$.\footnote{This results in $\epsilon$ ranging from $0.2$ to $0.08$, where the cube is scaled into $[-0.2,0.2]^3$.} 
One can observe linear complexity of all parts except the \textit{fft} that is constant for constant $n_q$. Note that the NFFT is linear in the number of nodes of the mesh.\\[0.1cm]
Next we compare our method for the case of uniform magnetization, i.e. $\boldsymbol{m} = (0,0,1)^T, \, M_s = 1$, in a sphere with radius $R$ and center at zero, where the exact solution is given as [$\boldsymbol{x} = (x^{(1)},x^{(2)},x^{(3)})$]
\begin{align}\label{phi_anal}
\phi^{int}(\boldsymbol{x}) = & \, \frac{x^{(3)}}{3},\\
\phi^{ext}(\boldsymbol{x}) = &\, R^3 \frac{x^{(3)}}{3 \left\|\boldsymbol{x}\right\|^3}, 
\end{align}
which can easily be verified by inserting into \eqref{scpot_classical}. We took the same parameters as in the first experiment. Fig.~\ref{error1} shows number of nodes versus the maximum of the point-wise absolute error at the nodes of the mesh, i.e. $l_\infty$-error, of the computed solution in $\overline{\Omega}$ compared to the analytical value.
One observes linear error decay.\\
We remark that $\phi_1^{int} \equiv 0$ [compare with \eqref{GR1}], since $\nabla\cdot\boldsymbol{m} \equiv 0$ in $\Omega$. Hence, this example only tests the computation of
$\phi_2$, cf. \eqref{GR2}.\\
In order to include the computation of $\phi_1$ in our tests, we take the example $\boldsymbol{m}(\boldsymbol{x}) = \boldsymbol{x}/\left\|\boldsymbol{x}\right\|$ in a sphere with radius $R$ and center at zero with exact solution
\begin{align}
 \phi^{int}(\boldsymbol{x}) & \,= \left\|\boldsymbol{x}\right\| - R,\\
 \phi^{ext}(\boldsymbol{x}) & \,= 0.
\end{align}
Since $\phi$ is zero at the boundary, we have $\phi^{int} = \phi_1^{int}$ in \eqref{GR1} and $[\frac{\partial\phi_2}{\partial{\boldsymbol{n}}}] = 0$, hence $\phi_2 = 0$ in \eqref{GR2}. 
Nevertheless, in our numerical test we also include the computation of $\phi_2$ through \eqref{slayer}. 
Tab.\ref{tab2} shows the errors in the $L_2(\Omega)$-norm, $H^1(\Omega)$-semi-norm and $H^1(\Omega)$-norm, i.e.
\begin{align}
 \left\|\phi - \phi_{\text{appr}} \right\|_{L^2(\Omega)} & \, = \Big(\int_{\Omega} (\phi - \phi_{\text{appr}})^2(\boldsymbol{x}) \,\text{d}\boldsymbol{x}\Big)^{1/2},\\
 |\phi - \phi_{\text{appr}} |_{H^1(\Omega)} & \, = \Big(\sum_{q=1}^3 \left\| \partial_q (\phi - \phi_{\text{appr}}) \right\|_{L^2(\Omega)}^2\Big)^{1/2},\\
 \left\|\phi - \phi_{\text{appr}} \right\|_{H^1(\Omega)} &\, = \Big(\left\|\phi - \phi_{\text{appr}} \right\|_{L^2(\Omega)}^2 + |\phi - \phi_{\text{appr}} |_{H^1(\Omega)}^2\Big)^{1/2},
\end{align}
respectively, which we calculated by taking the nodal interpolations of the exact and computed solutions. We took $n = 72, \, \alpha = 2$ and cut-off parameters $m=5$. Note that the $H^1(\Omega)$-semi-norm and thus the $H^1(\Omega)$-norm take the errors of the stray field $\boldsymbol{h}_s = - \nabla \phi$ into account.\\
\begin{table}
  \centering
  \begin{tabular}{c c c c c}
    \# elements & \# nodes  & $\left\|.\right\|_{L^2(\Omega)}$  & $|. |_{H^1(\Omega)}$ & $\left\|. \right\|_{H^1(\Omega)}$ \\ \hline\hline
    $3058$ & $678$   &  $1.0$e-$3$  &  $9.9$e-$3$ & $1.0$e-$2$  \\ \hline
    $7188$ & $1490$   & $8.9$e-$4$  &  $7.7$e-$3$ & $7.7$e-$3$  \\ \hline
    $14169$ & $2232$   & $7.2$e-$4$  &  $3.0$e-$3$ & $3.1$e-$3$  \\ \hline
   \end{tabular}
  \caption{Errors for magnetization $\boldsymbol{m}(\boldsymbol{x}) = \boldsymbol{x}/\left\|\boldsymbol{x}\right\|$ in a sphere with radius $0.2$ and center at zero measured in the $L^2(\Omega)$-norm, $H^1(\Omega)$-semi-norm and $H^1(\Omega)$-norm.}\label{tab2} 
\end{table}
\begin{table}
  \centering
  \begin{tabular}{c c c c c c c c c c}
    $N$    &   $M$  & $\epsilon_{nf}$ & $N_{nf}$ & $S^\prime$ & $n$    & energy        & error $[\%]$ & $t_{ff}$\,[sec] & $t_{nf}$\,[sec]\\ \hline\hline
    $1323$ & $1920$ & $0.15$          & $1323$   & $6$        & $48$   & $3.978$e-$02$ & $1.17$       & $1.0$e-$01$     & $1.9$e-$03$ \\ \hline
    \ditto & \ditto & $0.02$          & $1034$   & $8$        & \ditto & $3.825$e-$02$ & $4.97$       & $9.9$e-$02$     & $9.0$e-$04$ \\ \hline
    \ditto & \ditto & \ditto          & \ditto   & \ditto     & $72$   & \ditto        & \ditto       & $2.7$e-$01$     & $8.0$e-$04$ \\ \hline
    \ditto & \ditto & \ditto          & \ditto   & $9$        & \ditto & $3.928$e-$02$ & $2.41$       & $2.6$e-$01$     & $8.2$e-$04$ \\ \hline
    \ditto & \ditto & \ditto          & \ditto   & $10$       & \ditto & $3.996$e-$02$ & $1.47$       & $3.2$e-$01$     & $8.1$e-$04$ \\ \hline
    \ditto & \ditto & $0.05$          & \ditto   & \ditto     & \ditto & $3.977$e-$02$ & $1.19$       & $3.2$e-$01$     & $8.1$e-$04$ \\ \hline
    \ditto & \ditto & $0.01$          & $962$    & \ditto     & \ditto & $3.946$e-$02$ & $1.96$       & $2.9$e-$01$     & $7.3$e-$04$ \\ \hline
    $8405$ & $7680$ & $0.05$          & $8405$   & \ditto     & \ditto & $4.010$e-$02$ & $0.37$       & $3.3$e-$01$     & $6.7$e-$02$ \\ \hline
   \end{tabular}
  \caption{Errors of the stray field energy and timings for uniform magnetization in a $1\times 1 \times 0.1$ thin film (scaled into the $(-0.2,0.2)^3$-box) with respect to different parameters: $N$ number of nodes, $M$ number of surface elements, $\epsilon_{nf}$ near field radius, $N_{nf}$ number of nodes in the near field zone, 
  $S^\prime$ number of terms in the approximation of the kernel function by exponential sums, $n$ tensor grid size in one direction. The other quantities are: the stray field energy, the error of the stray field energy in $\%$, $t_{nf}$ time for near field part, $t_{ff}$ time for smooth part.}\label{tab3} 
\end{table}
Tab.~\ref{tab3} shows the errors of the stray field energy and corresponding timings for uniform magnetization in a $1\times 1 \times 0.1$ thin film (scaled into the $(-0.2,0.2)^3$-box) with respect to several different parameters. 
The stray field energy is calculated by using so-called \textit{mass lumping} for the computation of the stray field as described in (e.g.) \cite[Sec.~2.2.3]{schrefl_handbook}. The approximation is compared to the 'exact' reference value $4.025$e-$02$, which is taken from \cite{abert_2013}. 
In both non-uniform FFT directions the cut-off $m = 4$ was used.
\begin{table}
  \centering
  \begin{tabular}{c c c c c c c c c c}
    $N$    &   $M$  & $\epsilon_{nf}$ & $N_{nf}$ & $S^\prime$ & $n$    & energy        & error $[\%]$ & $t_{ff}$\,[sec] & $t_{nf}$\,[sec]\\ \hline\hline
    $1331$ & $1200$ & $0.15$          & $1304$   & $6$        & $48$   & $1.595$e-$01$ & $4.38$       & $1.4$e-$01$     & $2.9$e-$03$ \\ \hline
    \ditto & \ditto & $0.10$          & $1206$   & \ditto     & \ditto & \ditto        & \ditto       & $1.5$e-$01$     & $5.8$e-$03$ \\ \hline
    \ditto & \ditto & \ditto          & \ditto   & \ditto     & $72$   & \ditto        & \ditto       & $2.9$e-$01$     & $2.4$e-$03$ \\ \hline
    \ditto & \ditto & $0.08$          & $988$    & $8$        & \ditto & \ditto				 & \ditto       & $3.2$e-$01$     & $9.0$e-$04$ \\ \hline
    $2744$ & $2028$ & $0.10$          & $2528$   & $6$        & $48$   & $1.588$e-$01$ & $3.93$       & $1.6$e-$01$     & $3.8$e-$03$ \\ \hline
    \ditto & \ditto & $0.05$          & $1744$   & $8$        & $72$   & $1.589$e-$01$ & $3.99$       & $3.8$e-$01$     & $1.2$e-$03$ \\ \hline
    $9261$ & $4800$ & $0.10$          & $7930$   & \ditto     & \ditto & $1.570$e-$01$ & $2.75$       & $5.2$e-$01$     & $3.0$e-$02$ \\ \hline
   \end{tabular}
  \caption{Errors of the stray field energy and timings for flower-like magnetization in the unit cube (scaled into the $(-0.2,0.2)^3$-box) with respect to different parameters: $N$ number of nodes, $M$ number of surface elements, $\epsilon_{nf}$ near field radius, $N_{nf}$ number of nodes in the near field zone, 
  $S^\prime$ number of terms in the approximation of the kernel function by exponential sums, $n$ tensor grid size in one direction. The other quantities are: the stray field energy, the error of the stray field energy in $\%$, $t_{nf}$ time for near field part, $t_{ff}$ time for smooth part.}\label{tab4} 
\end{table}
Tab.~\ref{tab4} shows the errors of the stray field energy and corresponding timings for flower-like magnetization in the unit cube (scaled into the $(-0.2,0.2)^3$-box) with respect to several different parameters. The approximation is compared to the 'exact' reference value $1.528$e-$01$, which is taken from \cite{abert_2013}. 
In both non-uniform FFT directions the cut-off $m = 5$ was used.
\section{Errors, Complexity and Storage}\label{error_anal}
Remember that the approximation for the single layer potential is split into a near field correction and a smooth part, i.e.
\begin{align}\label{whole_scheme} 
\phi_2(\boldsymbol{x}_i) \approx \sum_{j = 1}^M g_{j} \int_{S_j} \, \mathcal{N}_{\text{NF}}( \boldsymbol{x}_i - \boldsymbol{y} ) \,\text{d}\sigma(\boldsymbol{y}) + \sum_{j = 1}^M g_j \int_{S_j} \, \mathcal{N}_s( \boldsymbol{x}_i - \boldsymbol{y} ) \,\text{d}\sigma(\boldsymbol{y}) =: \phi_{2}^{\text{NF}}(\boldsymbol{x}_i) + \phi_2^s(\boldsymbol{x}_i).
\end{align}
The scheme for the smooth part written in a compact form reads
\begin{align}\label{BEM_NFFT_compact2}
 \phi_2^s \approx \text{NFFT}\Big( \big(c_{\boldsymbol{l}}(\mathcal{N}_s)\big)_{\boldsymbol{l} \in I_{\boldsymbol{n}}} \odot \big(\text{FFT}(\mathcal{A})/c_{\boldsymbol{l}}(\widetilde{\Upsilon})\big)_{\boldsymbol{l}\in I_{\boldsymbol{n}}} \Big).
\end{align}
As pointed out in section~\ref{BEM-NFFT} and also numerically tested in section~\ref{window_fct}, the error that arises from approximating the tensor $\mathcal{B}$ with entries $b_{\boldsymbol{l}} = \sum_{j=1}^M g_{j}\,\int_{S_j} e^{-2\pi \mathrm{i} \boldsymbol{y}\cdot \boldsymbol{l}} \, \text{d}\sigma(\boldsymbol{y})$, i.e.
\begin{align}
\mathcal{B} \approx \big(\text{FFT}(\mathcal{A})/c_{\boldsymbol{l}}(\widetilde{\Upsilon})\big)_{\boldsymbol{l}\in I_{\boldsymbol{n}}},
\end{align}
behaves like that for the standard NFFT. The error bound in section~\ref{window_fct} shows that this error decays exponentially with increasing cut-off parameter $m$ and is independent of the tensor grid size $|I_{\boldsymbol{n}}|$.\\[0.1cm]
In order to be able to analyze the error dependence on $\boldsymbol{n}$ of the whole scheme \eqref{whole_scheme} one has to look at the kernel splitting in more detail, i.e.
\begin{align}
 \mathcal{N} = (\mathcal{N} - \mathcal{N}_s) + \mathcal{F}\mathcal{N}_s + (\mathcal{N}_s - \mathcal{F}\mathcal{N}_s).
\end{align}
In the scheme \eqref{whole_scheme} with \eqref{BEM_NFFT_compact2} for the smooth part, the contribution of $\mathcal{N}_s - \mathcal{F}\mathcal{N}_s$ is neglected. Thus, the error occurring from the approximation of the smooth kernel approximation $\mathcal{N}_s$ by its Fourier series approximation $\mathcal{F}\mathcal{N}_s$ has to be analyzed. 
Moreover, in order to get linear complexity in the near field correction, $(\mathcal{N} - \mathcal{N}_s)(\boldsymbol{x}) = 0$ is assumed for $\left\|\boldsymbol{x}\right\|> \epsilon$. Due to the approximation by exponential sums, compare with section~\ref{kernel_appr}, 
this yields a (uniform) error in the interval $[\epsilon,\beta]$, which is denoted as $E_{\text{NF}}$ in the following estimate. Overall, for the essential error arising in the summation in \eqref{whole_scheme} holds  
\begin{align}\label{error_est1}
 |\phi_{2}^{\text{NF}}(\boldsymbol{x}_i) + \phi_2^s(\boldsymbol{x}_i) - \big(\widetilde{\phi_{2}^{\text{NF}}}(\boldsymbol{x}_i) + \widetilde{\phi_2^s}(\boldsymbol{x}_i)\big)| \leq 
|\partial \Omega|\, \left\| \boldsymbol{g} \right\|_1 \,\Big(E_{\text{NF}} + \max_{\left\|\boldsymbol{x}\right\|< \tfrac{1}{2}} \left|\mathcal{N}_s(\boldsymbol{x}) - \mathcal{F}\mathcal{N}_s(\boldsymbol{x}) \right| \Big),
\end{align}
where $\widetilde{\phi_{2}^{\text{NF}}}(\boldsymbol{x}_i) + \widetilde{\phi_2^s}(\boldsymbol{x}_i)$ denotes the computed values and $\left\| \boldsymbol{g} \right\|_1 := \sum_{j = 1}^M | g_{j} |$.\\
Due to the tensor product structure of the Fourier coefficients of $\mathcal{N}_s$ (compare with section~\ref{kernel_appr}), also the Fourier series approximation has this structure, i.e.
\begin{align}
 \mathcal{F}\mathcal{N}_s(\boldsymbol{x}) = \,\sum_{k=1}^{S^\prime} \omega_k\,\mathcal{F}N_s^{(k)}(x_1) \,\mathcal{F}N_s^{(k)}(x_2)\, \mathcal{F}N_s^{(k)}(x_3).
\end{align}
It follows
\begin{equation}\label{error_est2}
\begin{aligned}
 \max_{\left\|\boldsymbol{x}\right\| < \tfrac{1}{2}} \left|\mathcal{N}_s(\boldsymbol{x}) - \mathcal{F}\mathcal{N}_s(\boldsymbol{x}) \right| & \,= \sum_{k=1}^{S^\prime} |\omega_k|  \max_{\left\|\boldsymbol{x}\right\|< \tfrac{1}{2}}| N_s^{(k)}(x_1) \,N_s^{(k)}(x_2)\, N_s^{(k)}(x_3) - \mathcal{F}N_s^{(k)}(x_1) \,\mathcal{F}N_s^{(k)}(x_2)\, \mathcal{F}N_s^{(k)}(x_3)|\\[0.1cm]
 & \,\leq \sum_{k=1}^{S^\prime} C_k\,|\omega_k|  \sum_{q=1}^3 \max_{\left\|\boldsymbol{x}\right\|< \tfrac{1}{2}}| N_s^{(k)}(x_q) - \mathcal{F}N_s^{(k)}(x_q) |,
\end{aligned}
\end{equation}
where an telescoping sum like $abc -\widetilde{a}\widetilde{b}\widetilde{c} = (a-\widetilde{a})bc + (b-\widetilde{b})\widetilde{a}c + (c-\widetilde{c})\widetilde{a}\widetilde{b}$ was used and 
$C_k$ is an upper bound for the products $bc,\widetilde{a}c$ and $\widetilde{a}\widetilde{b}$.\\
Adapting the proof of Th.~$3.4$ in \cite{potts2004fast} for the univariate case, the error $\max_{x_q < \tfrac{1}{2}}| N_s^{(k)}(x_q) - \mathcal{F}N_s^{(k)}(x_q) |$ for $N_s^{(k)}(x_q) = e^{-\gamma_k x_q^2}$ can be estimated by
\begin{align}\label{error_est3}
 \max_{x_q < \tfrac{1}{2}}| N_s^{(k)}(x_q) - \mathcal{F}N_s^{(k)}(x_q) | \leq A(\gamma_k,\eta_k^q) + B(\gamma_k,\eta_k^q),
\end{align}
where $\eta_k^q := \tfrac{\pi n_q}{2 \sqrt{\gamma_k}}$ and $A(\gamma_k,\eta_k^q) \sim e^{- (\eta_k^q)^2}$ and $B(\gamma_k,\eta_k^q) \sim e^{- \gamma_k/4}/\eta_k^q$.\\ 
The consequences of \eqref{error_est3} are twofold. First, for small $\gamma_k$ the term $B(\gamma_k,\eta_k^q)$ only gets small for large $n_q$, whereas for large $\gamma_k$ this term is negligible. In the first case (small $\gamma_k$) one can 
use boundary regularization or further scaling the domain $\Omega$ into, e.g., $(-0.2,0.2)^3$. This reduces the error $N_s^{(k)} - \mathcal{F}N_s^{(k)}$ in general, \cite{potts2004fast}.\\
On the other hand \eqref{error_est3} suggests to choose $n_q$ in the order of $\sqrt{\gamma_k}$, i.e. $n_q \sim \sqrt{\gamma_k}$, such that $\eta_k^{(q)} \geq 1$ and thus $A(\gamma_k,\eta_k^q)$ is small.\\
In order to get the relation between the tensor gird size $|I_{\boldsymbol{n}}|$ and $\epsilon$,\footnote{The left border of the interval of validity for the uniform approximation of the kernel cf. Sec.~\ref{kernel_appr}.} linear fitting of the quantities $\log \epsilon$ and $\log \gamma_{S^\prime}$, both, as functions of $S^\prime$, is used. This gives, for several precomputed lists from \cite{exp_sums_hackbusch}, the relation   
\begin{align}\label{para_est1}
 |I_{\boldsymbol{n}}| = \mathcal{O}(1/\epsilon^\delta),\quad \delta \approx 3.
\end{align}
Now, linear complexity of the near field computation requires that $\nu := \max|I_{\text{NF}_{\epsilon}}(S_j)|$ is much smaller than the total number of nodes, i.e. $N$, cf. section~\ref{near_field}. 
Assuming that the nodes near the boundary are reasonably uniformly distributed, means that the '$\epsilon-$balls' $I_{\text{NF}_{\epsilon}}(S_j)$ 
contain about the same number of nodes, namely $\nu$. If the even more idealistic assumption is made that the whole mesh is roughly uniform, then the volume of an $\epsilon$-ball is proportional to the ratio $\nu/N$, i.e. there should hold approximately $\epsilon \sim (\nu/N)^{1/3}$.\\
Together with \eqref{para_est1} this combines to 
\begin{align}\label{para_est2}
 |I_{\boldsymbol{n}}| = \mathcal{O}(N). 
\end{align}
Since the complexity of the proposed scheme for \eqref{whole_scheme} is $\mathcal{O}\big(M+N + (\prod_{q=1}^3 n_q) (\log  \prod_{q=1}^3 n_q)\big)$, the assumption of a roughly uniform mesh, 
together with the error investigation above, gives rise to the scaling 
\begin{align}
\mathcal{O}( M + N + N \log N).  
\end{align}
Concerning the storage requirements of the method, one has to collect all pieces in the setup of Alg.~\ref{alg_tot}. The storage of the LU decomposition of the stiffness matrix can be of quadratic order in the number of mesh nodes $N$. Nevertheless, a preconditioned iterative procedure would need linear storage in $N$.\\ As indicated in Sec.~\ref{Gridding}, the storage for the sparse matrix $\boldsymbol{A}$ is $\mu M$, where $M$ is the number of surface elements and $\mu$ is of the order of $m^3$, the cubed cut-off parameter. Tensor train compression can reduce the constant $\mu$.\\ Due to the approximation with exponential sums, the Fourier coefficients of the multivariate function $\mathcal{N}_s$ only require $S^\prime \sum_{q=1}^3 n_q$ complex numbers, instead of $\prod_{q=1}^{3} n_q$, if one can not rely on a separable structure, cf. Sec.~\ref{kernel_appr}. Similar, the Fourier coefficients of the window functions (tensor product of univariate functions) is only stored for the one-dimensional case, yielding a storage requirement of $\sum_{q=1}^3 n_q$. Finally, the near field integrals have to be stored, which is at most $\nu M,\, \nu = \max_j |I_{{NF}_\epsilon}(S_j)|$ numbers. Overall, the storage requirements\footnote{A preconditioned iterative procedure instead of LU decomposition is assumed.} are
\begin{align}
\mathcal{O}\big(N + (\mu +\nu) M + (S^\prime + 1) \sum_{q=1}^3 n_q\big),
\end{align}   
which is linear in $N, M$ and the $n_q$'s. 
\section*{Conclusion}
We introduced a P$1$ finite element method for the computation of the micromagnetic scalar potential based on the ansatz of Garc\'{i}a-Cervera and Roma. 
The potential is computed by a splitting $\phi = \phi_1 + \phi_2$, where $\phi_1$ is solved by a Poisson equation with zero Dirichlet boundary conditions and $\phi_2$ by evaluation of the single layer potential.
Our contribution is the development of a method to compute the single layer potential at all nodes of a tetrahedral mesh in linear time by means of Fourier approximation of a smoothed kernel and near field correction.\\
The discretized integral operator splits into a part with smooth and singular kernel. The latter one has small support and therefore allows a computation by sparse summation, while for the smooth part Fourier techniques can be applied.
Due to the unstructured FE-mesh, generalizations of discrete Fourier transforms arise, where we developed efficient implementations.\\
Overall the method scales quasi linearly in the number of nodes and linearly in the number of surface elements. Similar, the storage requirements are linear in the number of surface elements, where we further introduced tensor train compression to reduce the constant in the 
storage estimate for the gridding procedure.\\
We used exponential sums to obtain an entirely smooth and separable approximation of the Fourier coefficients of the Newtonian potential. As a consequence of the above mentioned splitting, which includes a near field correction, the only essential error of our method (within the P$1$ FEM framework) is due to this approximation, cf. eqn.~\eqref{splitting2}. 
Nevertheless, numerical experiments for test cases with known analytical solutions show accurate approximations.\\  
Future work might include detailed mathematical analysis of this error, as well as possible extension to higher order finite element and boundary element methods.\\ 
Finally, we stress that our approach could be seen as a generalization to general geometries of the FFT techniques used in finite difference based micromagnetic methods for cuboid and equispaced computational domains, since our scheme shows some analogy to the convolution theorem which is used there, cf. eqn.~\eqref{convtheo}.
\section*{Acknowledgments} 
The first author wants to thank Stefan Schnabl and Michael Srb for helpful discussions on this topic.\\
Financial support by the Austrian Science Fund (FWF) SFB ViCoM (F4112-N13) is gratefully acknowledged.
%

\begin{thebibliography}{32}
\expandafter\ifx\csname natexlab\endcsname\relax\def\natexlab#1{#1}\fi
\providecommand{\bibinfo}[2]{#2}
\ifx\xfnm\relax \def\xfnm[#1]{\unskip,\space#1}\fi
\bibitem[{Fidler and Schrefl(2000)}]{fidler_2000}
\bibinfo{author}{J.~Fidler}, \bibinfo{author}{T.~Schrefl}, \bibinfo{journal}{J.
  Phys. D: Appl. Phys.} \bibinfo{volume}{33} (\bibinfo{year}{2000}).
\bibitem[{d'Aquino et~al.(2005)d'Aquino, Serpico, and
  Miano}]{d2005geometrical}
\bibinfo{author}{M.~d'Aquino}, \bibinfo{author}{C.~Serpico},
  \bibinfo{author}{G.~Miano}, \bibinfo{journal}{Journal of Computational
  Physics} \bibinfo{volume}{209} (\bibinfo{year}{2005})
  \bibinfo{pages}{730--753}.
\bibitem[{Jackson(1999)}]{jackson_classical_1999}
\bibinfo{author}{J.~D. Jackson}, \bibinfo{journal}{Am. J. Phys}
  \bibinfo{volume}{67} (\bibinfo{year}{1999}).
\bibitem[{Engel et~al.(2007)Engel, Rashba, and Halperin}]{engel2007handbook}
\bibinfo{author}{H.-A. Engel}, \bibinfo{author}{E.~I. Rashba},
  \bibinfo{author}{B.~I. Halperin}, \bibinfo{title}{Handbook of Magnetism and
  Advanced Magnetic Materials}, \bibinfo{publisher}{John Wiley \& Sons Ltd.,
  Chichester, UK}, \bibinfo{year}{2007}.
\bibitem[{Aurada et~al.(2012)Aurada, Feischl, F{\"u}hrer, Karkulik, Melenk, and
  Praetorius}]{aurada2012classical}
\bibinfo{author}{M.~Aurada}, \bibinfo{author}{M.~Feischl},
  \bibinfo{author}{T.~F{\"u}hrer}, \bibinfo{author}{M.~Karkulik},
  \bibinfo{author}{J.~M. Melenk}, \bibinfo{author}{D.~Praetorius},
  \bibinfo{journal}{Computational Mechanics}  (\bibinfo{year}{2012})
  \bibinfo{pages}{1--21}.
\bibitem[{Carstensen and Stephan(1995)}]{carstensen1995adaptive}
\bibinfo{author}{C.~Carstensen}, \bibinfo{author}{E.~P. Stephan},
  \bibinfo{journal}{ESAIM: Mathematical Modelling and Numerical
  Analysis-Mod{\'e}lisation Math{\'e}matique et Analyse Num{\'e}rique}
  \bibinfo{volume}{29} (\bibinfo{year}{1995}) \bibinfo{pages}{779--817}.
\bibitem[{Abert et~al.(2012)Abert, Exl, Selke, Drews, and Schrefl}]{abert_2013}
\bibinfo{author}{C.~Abert}, \bibinfo{author}{L.~Exl},
  \bibinfo{author}{G.~Selke}, \bibinfo{author}{A.~Drews},
  \bibinfo{author}{T.~Schrefl}, \bibinfo{journal}{Journal of Magnetism and
  Magnetic Materials} \bibinfo{volume}{326} (\bibinfo{year}{2012})
  \bibinfo{pages}{176--185}.
\bibitem[{Donahue and McMichael(2007)}]{donahue_2007}
\bibinfo{author}{M.~J. Donahue}, \bibinfo{author}{R.~D. McMichael},
  \bibinfo{journal}{IEEE Trans. Magn.} \bibinfo{volume}{43}
  (\bibinfo{year}{2007}) \bibinfo{pages}{2878 -- 2880}.
\bibitem[{Blue and Scheinfein(1991)}]{blue_1991}
\bibinfo{author}{J.~Blue}, \bibinfo{author}{M.~Scheinfein},
  \bibinfo{journal}{IEEE Trans. Magn.} \bibinfo{volume}{27}
  (\bibinfo{year}{1991}) \bibinfo{pages}{4778 --4780}.
\bibitem[{Long et~al.(2006)Long, Ong, Liu, and Li}]{long_2006}
\bibinfo{author}{H.~Long}, \bibinfo{author}{E.~Ong}, \bibinfo{author}{Z.~Liu},
  \bibinfo{author}{E.~Li}, \bibinfo{journal}{IEEE Trans. Magn.}
  \bibinfo{volume}{42} (\bibinfo{year}{2006}) \bibinfo{pages}{295 -- 300}.
\bibitem[{Van~de Wiele et~al.(2008)Van~de Wiele, Olyslager, and
  Dupr{\'e}}]{van2008application}
\bibinfo{author}{B.~Van~de Wiele}, \bibinfo{author}{F.~Olyslager},
  \bibinfo{author}{L.~Dupr{\'e}}, \bibinfo{journal}{Journal of Computational
  Physics} \bibinfo{volume}{227} (\bibinfo{year}{2008})
  \bibinfo{pages}{9913--9932}.
\bibitem[{Livshitz(2009)}]{livshitz}
\bibinfo{author}{B. Livshitz}, \bibinfo{author}{A. Boag}, \bibinfo{author}{H.~N. Bertram}, 
\bibinfo{author}{V. Lomakin},
  \bibinfo{journal}{Journal of Applied Physics} \bibinfo{volume}{105}
  (\bibinfo{year}{2009}) \bibinfo{pages}{07D541}.  
\bibitem[{Chang(2009)}]{chang}
\bibinfo{author}{R. Chang}, \bibinfo{author}{S. Li}, \bibinfo{author}{M.~V. Livshitz}, 
\bibinfo{author}{V. Lomakin},
  \bibinfo{journal}{Journal of Applied Physics} \bibinfo{volume}{109}
  (\bibinfo{year}{2011}) \bibinfo{pages}{07D358}.  
\bibitem[{Goncharov et~al.(2010)Goncharov, Hrkac, Dean, and
  Schrefl}]{goncharov_2010}
\bibinfo{author}{A.~Goncharov}, \bibinfo{author}{G.~Hrkac},
  \bibinfo{author}{J.~Dean}, \bibinfo{author}{T.~Schrefl},
  \bibinfo{journal}{Journal of Computational Physics} \bibinfo{volume}{229}
  (\bibinfo{year}{2010}) \bibinfo{pages}{2544--2549}.
\bibitem[{Exl et~al.(2012{\natexlab{a}})Exl, Auzinger, Bance, Gusenbauer,
  Reichel, and Schrefl}]{exl_fast_2012_2}
\bibinfo{author}{L.~Exl}, \bibinfo{author}{W.~Auzinger},
  \bibinfo{author}{S.~Bance}, \bibinfo{author}{M.~Gusenbauer},
  \bibinfo{author}{F.~Reichel}, \bibinfo{author}{T.~Schrefl},
  \bibinfo{journal}{J. Comput. Phys.} \bibinfo{volume}{231}
  (\bibinfo{year}{2012}{\natexlab{a}}) \bibinfo{pages}{2840--2850}.
\bibitem[{Exl et~al.(2012{\natexlab{b}})Exl, Abert, Mauser, Schrefl, Stimming,
  and Suess}]{exl2012fft}
\bibinfo{author}{L.~Exl}, \bibinfo{author}{C.~Abert}, \bibinfo{author}{N.~J.
  Mauser}, \bibinfo{author}{T.~Schrefl}, \bibinfo{author}{H.~P. Stimming},
  \bibinfo{author}{D.~Suess}, \bibinfo{journal}{Math. Models Methods Appl. Sci. (in print),  doi: 10.1142/S0218202514500109}
   (\bibinfo{year}{2014}{\natexlab{b}}).
\bibitem[{phillips(1997)precorrected}]{phillips}
\bibinfo{author}{J.~R.~Phillips}, \bibinfo{author}{J.~K.~Jacob},
  \bibinfo{journal}{Computer-Aided Design of Integrated Circuits and Systems, IEEE Transactions on},
  \bibinfo{volume}{16} (\bibinfo{year}{1997})
  \bibinfo{pages}{1059--1072}.   
\bibitem[{Keiner et~al.(2009)Keiner, Kunis, and Potts}]{keiner2009using}
\bibinfo{author}{J.~Keiner}, \bibinfo{author}{S.~Kunis},
  \bibinfo{author}{D.~Potts}, \bibinfo{journal}{ACM Transactions on
  Mathematical Software (TOMS)} \bibinfo{volume}{36} (\bibinfo{year}{2009})
  \bibinfo{pages}{19}.
\bibitem[{Kritsikis et~al.(2008)Kritsikis, Toussaint, Fruchart, Szambolics, and
  Buda-Prejbeanu}]{kritsikis2008fast}
\bibinfo{author}{E.~Kritsikis}, \bibinfo{author}{J.-C. Toussaint},
  \bibinfo{author}{O.~Fruchart}, \bibinfo{author}{H.~Szambolics},
  \bibinfo{author}{L.~Buda-Prejbeanu}, \bibinfo{journal}{Applied Physics
  Letters} \bibinfo{volume}{93} (\bibinfo{year}{2008})
  \bibinfo{pages}{132508--132508}.
\bibitem[{Schrefl et~al.(2007)}]{schrefl_handbook}
\bibinfo{author}{T.~Schrefl}, \bibinfo{author}{G.~Hrkac}, \bibinfo{author}{S.~Bance}, 
\bibinfo{author}{D.~Suess}, \bibinfo{author}{O.~Ertl}, \bibinfo{author}{J.~Fidler}, 
\bibinfo{title}{Numerical methods in micromagnetics (finite element method)},
\bibinfo{publisher}{John Wiley \& Sons Ltd., Chichester, UK} (\bibinfo{year}{2007}).
\bibitem[{Potts and Steidl(2003)}]{potts2003fast}
\bibinfo{author}{D.~Potts}, \bibinfo{author}{G.~Steidl}, \bibinfo{journal}{SIAM
  Journal on Scientific Computing} \bibinfo{volume}{24} (\bibinfo{year}{2003})
  \bibinfo{pages}{2013--2037}.
\bibitem[{Brunotte et~al.(1992)Brunotte, Meunier, and Imhoff}]{brunotte_1992}
\bibinfo{author}{X.~Brunotte}, \bibinfo{author}{G.~Meunier},
  \bibinfo{author}{J.~Imhoff}, \bibinfo{journal}{IEEE Trans. Magn.}
  \bibinfo{volume}{28} (\bibinfo{year}{1992}) \bibinfo{pages}{1663 --1666}.
\bibitem[{Fredkin and Koehler(1990)}]{fredkin_1990}
\bibinfo{author}{D.~Fredkin}, \bibinfo{author}{T.~Koehler},
  \bibinfo{journal}{IEEE Trans. Magn.} \bibinfo{volume}{26}
  (\bibinfo{year}{1990}) \bibinfo{pages}{415 --417}.
\bibitem[{Knittel et~al.(2009)Knittel, Franchin, Bordignon, Fischbacher,
  Bending, and Fangohr}]{knittel_2009}
\bibinfo{author}{A.~Knittel}, \bibinfo{author}{M.~Franchin},
  \bibinfo{author}{G.~Bordignon}, \bibinfo{author}{T.~Fischbacher},
  \bibinfo{author}{S.~Bending}, \bibinfo{author}{H.~Fangohr},
  \bibinfo{journal}{J. Appl. Phys.} \bibinfo{volume}{105}
  (\bibinfo{year}{2009}) \bibinfo{pages}{07D542}.
\bibitem[{Garcia-Cervera and Roma(2006)}]{garcia2006adaptive}
\bibinfo{author}{C.~J. Garcia-Cervera}, \bibinfo{author}{A.~M. Roma},
  \bibinfo{journal}{Magnetics, IEEE Transactions on} \bibinfo{volume}{42}
  (\bibinfo{year}{2006}) \bibinfo{pages}{1648--1654}.
\bibitem[{Dutt and Rokhlin(1993)}]{dutt1993fast}
\bibinfo{author}{A.~Dutt}, \bibinfo{author}{V.~Rokhlin}, \bibinfo{journal}{SIAM
  Journal on Scientific computing} \bibinfo{volume}{14} (\bibinfo{year}{1993})
  \bibinfo{pages}{1368--1393}.
\bibitem[{Beylkin(1995)}]{beylkin1995fast}
\bibinfo{author}{G.~Beylkin}, \bibinfo{journal}{Applied and Computational
  Harmonic Analysis} \bibinfo{volume}{2} (\bibinfo{year}{1995})
  \bibinfo{pages}{363--381}.
\bibitem[{Potts et~al.(2001)Potts, Steidl, and Tasche}]{potts2001fast}
\bibinfo{author}{D.~Potts}, \bibinfo{author}{G.~Steidl},
  \bibinfo{author}{M.~Tasche}, in: \bibinfo{booktitle}{Modern sampling theory},
  \bibinfo{publisher}{Springer}, \bibinfo{year}{2001}, pp.
  \bibinfo{pages}{247--270}.
\bibitem[{Greengard et~al.(2004)}]{greengard_nufft}
\bibinfo{author}{L.~Greengard}, \bibinfo{author}{J.-Y.~Lee},
  \bibinfo{journal}{SIAM review}
  \bibinfo{volume}{46} (\bibinfo{year}{2004}) \bibinfo{pages}{443--454}.   
\bibitem[{Funken et~al.(2011)Funken, Praetorius, and
  Wissgott}]{funken2011efficient}
\bibinfo{author}{S.~Funken}, \bibinfo{author}{D.~Praetorius},
  \bibinfo{author}{P.~Wissgott}, \bibinfo{journal}{Comput. Methods Appl. Math.}
  \bibinfo{volume}{11} (\bibinfo{year}{2011}) \bibinfo{pages}{460--490}.
\bibitem[{Oseledets and Tyrtyshnikov(2010)}]{tt_tensor2}
\bibinfo{author}{I.~Oseledets}, \bibinfo{author}{E.~Tyrtyshnikov},
  \bibinfo{journal}{Linear Algebra and its Applications} \bibinfo{volume}{432}
  (\bibinfo{year}{2010}).
\bibitem[{Elbel and Steidl(1998)}]{elbel_1998}
\bibinfo{author}{A.~Elbel}, \bibinfo{author}{G.~Steidl}, in:
  \bibinfo{editor}{C.~Chui}, \bibinfo{editor}{L.~Schumaker} (Eds.),
  \bibinfo{booktitle}{In: Approximation Theory IX,},
  \bibinfo{address}{Vanderbuilt University Press,}, pp. \bibinfo{pages}{39
  --46}.
\bibitem[{Potts(2003)}]{potts_habil}
\bibinfo{author}{D.~Potts}, \bibinfo{publisher}{Habilitationsschrift,
  Universit\"at zu L\"ubeck}, \bibinfo{year}{2003}.
\bibitem[{Braess and Hackbusch(2009)}]{braess2009efficient}
\bibinfo{author}{D.~Braess}, \bibinfo{author}{W.~Hackbusch}, in:
  \bibinfo{booktitle}{Multiscale, nonlinear and adaptive approximation},
  \bibinfo{publisher}{Springer}, \bibinfo{year}{2009}, pp.
  \bibinfo{pages}{39--74}.
\bibitem[{exp(????)}]{exp_sums_hackbusch}
\bibinfo{url}{\url{http://www.mis.mpg.de/scicomp/EXP_SUM/1_sqrtx/}}.
\bibitem[{Lyness and Cools(1994)}]{lyness1994survey}
\bibinfo{author}{J.~N. Lyness}, \bibinfo{author}{R.~Cools},
  \bibinfo{journal}{Proc. Symposia Appl. Math} \bibinfo{volume}{48}
  (\bibinfo{year}{1994}) \bibinfo{pages}{127--150}.
\bibitem[{Potts et~al.(2004)Potts, Steidl, and Nieslony}]{potts2004fast}
\bibinfo{author}{D.~Potts}, \bibinfo{author}{G.~Steidl},
  \bibinfo{author}{A.~Nieslony}, \bibinfo{journal}{Numerische Mathematik}
  \bibinfo{volume}{98} (\bibinfo{year}{2004}) \bibinfo{pages}{329--351}.
 

\end{thebibliography}

\end{document}